\documentclass[]{JHEP3}
\usepackage{cite}
\usepackage{pinlabel}
\usepackage{graphicx}
\usepackage{amsmath}
\usepackage{subfigure,caption}

\title{Electroweak gauge boson polarisation at the LHC}
\author{W.J.~Stirling, E.~Vryonidou\\Cavendish Laboratory, J.J. Thomson Avenue,
Cambridge CB3 0HE, UK}
\preprint{Cavendish-HEP-2012/06}

\abstract
{We study the polarisation of gauge bosons produced at the LHC. Polarisation
effects for
$W$ bosons manifest themselves in the angular distributions of the lepton and in the distributions of lepton transverse momentum and missing transverse
energy. The distributions also depend on the selection
cuts, with kinematic effects competing with polarisation effects. The polarisation is discussed for a range
of different processes producing $W$ bosons: $W+$jets, $W$ from top (single and pair)
production, $W$ pair production and $W$ production in association with a $Z$ or
Higgs boson. The relative contributions of the different polarisation states varies from process to process, 
reflecting the dynamics of the underlying hard-scattering process. 
We also present results for the polarisation of the $Z$ boson produced in association with QCD jets at the LHC, 
and comment on the differences between $W$ and $Z$ production.}

\begin{document}
\tableofcontents
\section{Introduction}
The production of vector bosons has been extensively studied at past and present high-energy collider
experiments. Measurements of processes involving weak vector bosons are important both for 
confirming Standard Model (SM) electroweak predictions and in the search for evidence of New Physics. 
Recent LHC analyses have studied many different processes involving $W$ and $Z$ bosons. 
These include $W$ and $Z$ production both in association with QCD jets and in purely electroweak processes.

In the search for New Physics it is important to be able to accurately predict the corresponding SM backgrounds. 
In this context, $W,Z$+ jets production has been extensively studied in the
literature, with NLO corrections recently calculated for up to four associated jets \cite{Berger:2010zx,Ita:2011wn}. 
Apart from precise knowledge of the SM prediction, another important tool in the search for New Physics is
 the use of underlying properties to distinguish potential signals from the SM background. 
One such property is the {\it polarisation} of gauge bosons and the 
resulting distributions of lepton transverse momentum and missing transverse energy. $W$ bosons produced along the 
beam direction have long been known to be predominantly left-handed at the LHC~\cite{Ellis:1991qj}.
 In the study of Ref.~\cite{Berger:2009ep},  which considered the lepton transverse momentum and missing energy distributions, 
it was observed that $W$ bosons produced at the LHC in association with QCD jets are in general preferentially left-handed. 
The underlying physics leading to this observation has been explored in
\cite{Bern:2011ie} for $W+1$~jet, where the polarisation of $W$ bosons at non-zero transverse momentum, 
the angular distributions of the final-state decay leptons 
and the corresponding angular coefficients have been considered in detail. 
The dependence of the polarisation on the number of jets and the NLO pQCD
corrections has also been investigated, with the results shown to be rather stable. 

The CMS collaboration at the LHC has measured \cite{Chatrchyan:2011ig} the
polarisation of $W$ bosons produced in association with QCD jets at large transverse momentum, 
and demonstrated good agreement with the SM predictions presented in \cite{Bern:2011ie}. ATLAS has also very recently reported
 the measurement of the $W$ polarisation using the 2010 LHC data set in \cite{Aad:2012nn}.
In addition to the polarisation fractions of $W$ bosons, the  angular distributions of the produced leptons  
and the corresponding angular coefficients have also been investigated experimentally. 
An earlier analysis of the Tevatron ($p\bar p$) data has
 extracted the angular distribution of leptons from $W$ decays \cite{Acosta:2005dn} and the corresponding angular 
coefficients. Similarly, the full spin density matrix of the $W$ boson has been investigated in a series of
 phenomenological studies, see for example \cite{Collins:1977iv,Lam:1980uc,Mirkes:1992hu,Mirkes:1994eb}.

Measurements of the polarisation of $W$ bosons from other processes have also been undertaken by 
high-energy collider experiments. The polarisation of $W$ bosons from top pair production and decay 
has been investigated in the literature and measured by the Tevatron 
experiments \cite{Abazov:2007ve,Abazov:2010jn,Aaltonen:2008ei,Aaltonen:2010ha}. 
In this case the polarisation of the positively charged $W$ boson in the top rest frame is found 
to be predominantly longitudinal. The results are consistent with the SM predictions and therefore measurements of the $W$ 
polarisation in top pair production have also been employed to set limits on anomalous $Wtb$ 
couplings~\cite{AguilarSaavedra:2006fy}. The polarisation of weak bosons produced in pairs has 
been discussed in \cite{Bilchak:1984gv,Duncan:1985vj,Willenbrock:1987xz} and measurements of the $W$ 
polarisation have been performed by LEP experiments for $W$ pair production in \cite{Achard:2002bv,Abbiendi:2003wv}. 
The results were used to set limits on anomalous triple gauge boson couplings in \cite{Abdallah:2008sf}.

At the LHC, further exploration of the polarisation of gauge bosons will provide useful insight 
on their production mechanisms in kinematic regions not previously accessible at other colliders. 
This will prove useful both as a confirmation of the SM predictions but more importantly in the search 
for new interactions which can lead to different polarisation behaviour. 
In the absence of any deviation from the SM, measurements can be used to set more stringent limits 
on anomalous couplings, similarly to the Tevatron and LEP studies.
  
 In this study we revisit the polarisation of $W$ bosons produced in association with QCD jets in 
high-energy hadron-hadron collisions: how this is defined and computed and how it affects the shapes 
of the observable (lepton and missing energy) distributions. In Section~2
 we also investigate how the angular distributions of the decay leptons is affected by the introduction of
 selection cuts on the leptons and jets. In Section~3 we extend the analysis 
 to $W$ bosons from top pair production and decay, and then in Section~4 to a number of 
other processes in which $W$ bosons are produced. In Section~5 we present corresponding results
for the $Z$ boson, commenting on the differences between the two weak bosons, and we present our conclusions in Section~6.

\section{Polarisation in $W +$~jets production}
\label{sec:2}
\subsection{$W  +$~one jet production}
The observation that led the authors of \cite{Bern:2011ie} to study the
polarisation of $W$ bosons at large transverse momentum was the characteristic shape of distributions of the $W$ decay products.
The charge asymmetry ratio of $W$ and lepton $p_T$ distributions is strongly influenced by polarisation effects both in the $W$ 
boson production and decay.\footnote{In this paper we will focus mainly on differences in the distributions of the $W$ 
and lepton transverse momenta due to polarisation, but effects of the same origin can be observed in the corresponding 
rapidity distributions.} In this section we begin by considering $W+1$~jet production to explore the $W$ polarisation effects. 
The processes involved in $W+1$~jet production provide a simpler underlying mechanism and give a more direct handle on the 
kinematics. Results for more than one jet will be presented in subsequent sections.

\begin{figure}[h]
 \begin{minipage}[b]{0.5\linewidth}
\centering
\includegraphics[scale=0.6]{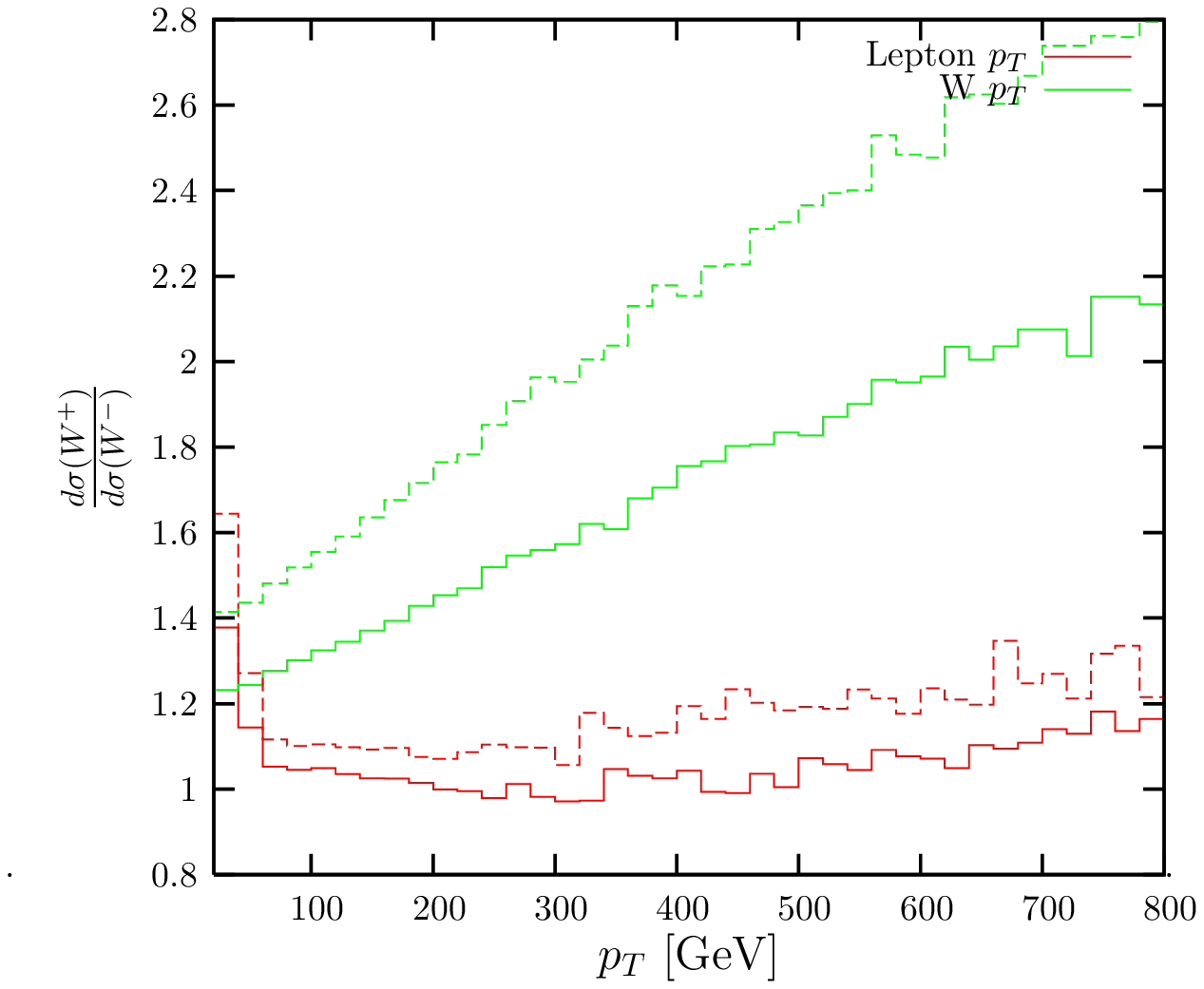}
\caption{Ratio of differential distributions for $W^+$ and $W^-$ for $W+1$~jet. Dashed: 7~TeV, solid: 14~TeV. 
Cuts: lepton, jet, missing $p_T>20$~GeV
and $|\eta|<2.5$ for both leptons and jets. }
\label{WpWm}
\end{minipage}
\hspace{0.5cm}
 \begin{minipage}[b]{0.5\linewidth}
 \centering
 \includegraphics[trim=1.3cm 0 0 0,scale=0.6]{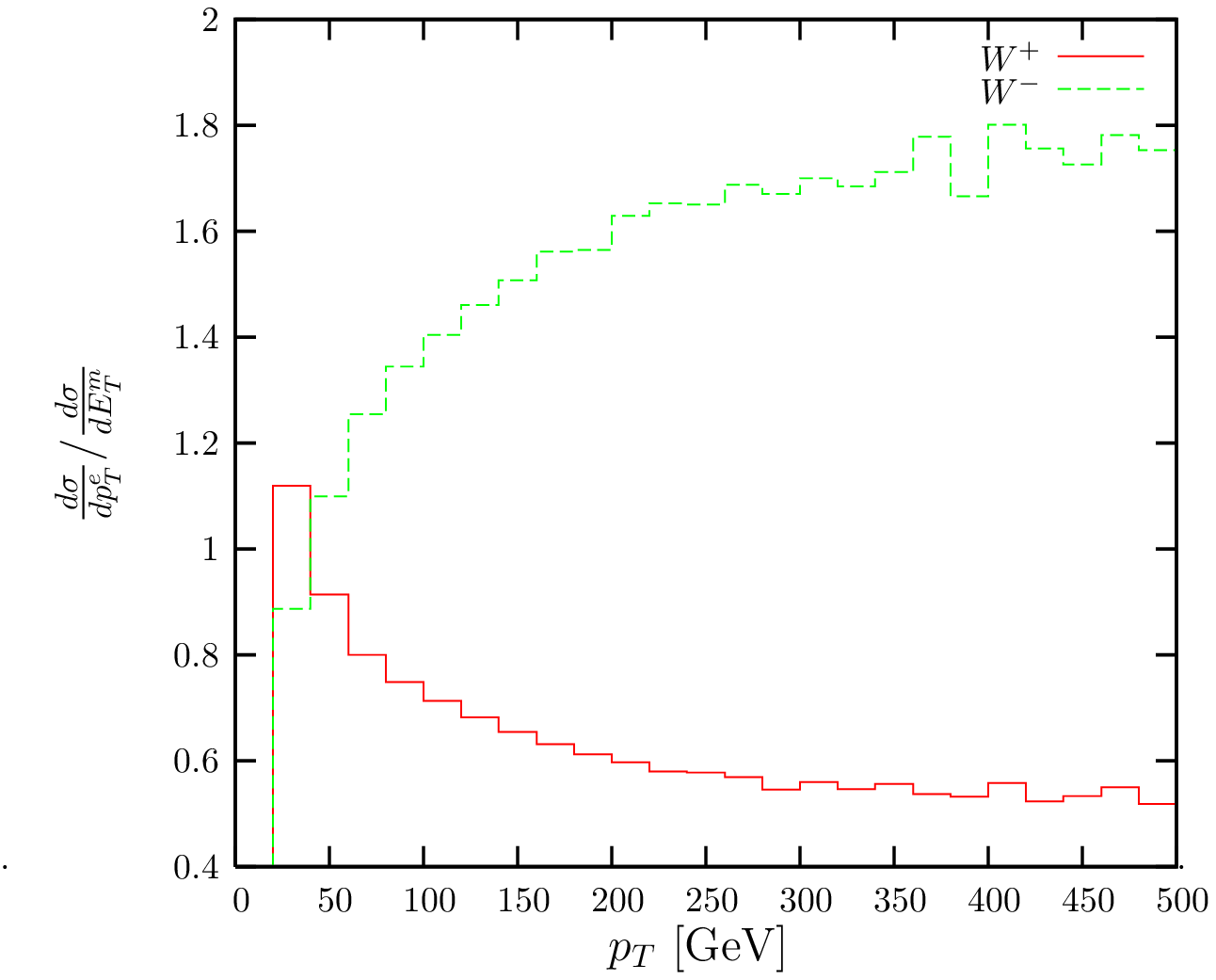}
\caption{Ratio of the differential distribution of the lepton transverse momentum to the distribution of the 
missing transverse energy for both $W^+$ and $W^-$ at 7~TeV. Imposed cuts as in Fig.~1.}
\label{lptmet}
 \end{minipage}
\end{figure}

 The ratio of distributions for $W+1$~jet production is
shown in Fig.~\ref{WpWm} for 7 and 14 TeV proton-proton collisions, for a standard set of final-state cuts appropriate
for $W$ studies as listed in the figure caption, and with renormalisation and factorisation scales set to $M_W$. 
This result and all subsequent quantitative results in this work have been obtained using LO cross sections and 
MSTW2008LO PDFs \cite{Martin:2009iq}. We will comment on the influence of next-to-leading order (NLO) corrections below.

The asymmetry in the W$^\pm$  $p_T$ distributions is manifest in Fig.~\ref{WpWm}, with the ratio of 
$W^+$ and $W^-$ $p_T^W$ distributions increasing monotonically with increasing $W$ transverse momentum. 
This is simply due to valence quarks becoming more important at high subprocess energies 
(equivalently, large momentum fractions $x$). At large $x$ the dominance of the valence 
$u-$quark parton distribution over that of the $d-$quark leads to an increasing ratio of $W^+$ over $W^-$. 
The effect is stronger at 7 TeV than at 14 TeV, since the corresponding $x$ values are larger for the same 
$p_T^W$.  This asymmetry between $W^+$ and $W^-$ at the LHC can also be used as a diagnostic tool for the 
presence of New Physics, see \cite{Kom:2010mv}. 

The difference between the electron and positron  $p_T$ distributions is characteristic of the dominant $W$ 
left-handed polarisation. In contrast to the $W^+/W^-$ ratio,  the ratio of the corresponding lepton
transverse momentum distributions {\em decreases} from small to medium $p_T$. The small increase at very 
high $p_T$ is again related to the relative strength of valence quark PDFs at high $x$. In contrast, the corresponding 
ratio for the missing transverse energy is an {\em increasing} function across the whole $p_T$ range. 
An effect of the same origin is observed in the characteristic shape of the ratio of the charged lepton $p_T$ distribution 
to the missing $E_T$ distribution, where the polarisation properties translate into a decreasing ratio for 
$W^+$ and an increasing ratio for $W^-$, as shown in Fig.~\ref{lptmet} for the LHC at 7~TeV. As already noted in \cite{Bern:2011ie}, 
similar shapes are obtained when more jets are present. We note here that all following plots presented in this paper are obtained for the LHC at 7~TeV.

\begin{figure}[h] 
\centering
\includegraphics[scale=0.6]{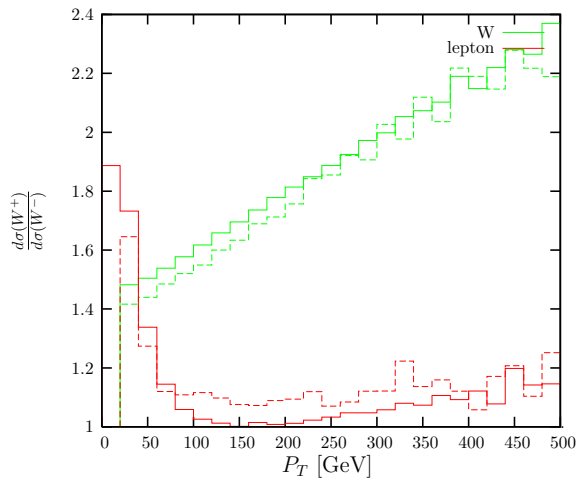}
\caption{Effect of cuts on the shape of the differential distributions at 7~TeV: only a jet minimum $p_T$ cut of 
30~GeV (solid) and for the full set of cuts as in Fig.~1 (dashed).}
\label{Wp730}
\end{figure}

If we relax the cuts so that the only cut is on the jet $p_T > 30$~GeV (as in \cite{Bern:2011ie}), then the result 
at 7~TeV is shown in Fig.~\ref{Wp730}, {\it cf.}  Fig.~\ref{WpWm} in which a full set of cuts have been
imposed. Evidently the basic shape characteristics are still present with the
introduction of cuts having little impact at large transverse momenta. 

The polarisation fractions of the $W$ boson in the helicity frame corresponding to the distributions shown in Fig.~\ref{Wp730} have
been obtained in \cite{Bern:2011ie}. In general, the angular distribution of the $W^+$ decay products in the $W^+$ rest 
frame is described by:
\begin{equation}
 \frac{1}{\sigma}\frac{d\sigma}{d\text{cos}\theta^*}=\frac{3}{8}(1-\text{cos}\theta^*)^2 f_L+\frac{3}{8}(1+\text{cos}\theta^*)^2f_R+\frac{3}{4}\text{sin}^2\theta^*f_0,
\label{diffeq}
\end{equation}where $\theta^*$ is the angle in the $W$ rest frame between the charged lepton and the $W$ flight direction in the lab frame, and $f_{0,L,R}$ are the polarisation fractions. For $W^-$, $f_R$ and $f_L$ are interchanged.
The normalisation is chosen so that $f_0+f_L+f_R=1$ and any dependence on the azimuthal angle has been integrated out.
In \cite{Bern:2011ie} it is noted that $\sigma$ in Eq.~(\ref{diffeq}) can be any differential cross section that does not depend 
on the kinematics of the individual leptons. In the literature the definition of $\theta^*$ varies between 
different studies, with the Collins-Soper frame \cite{Collins:1977iv} definition being used extensively.

Based on the structure defined in Eq.~(\ref{diffeq}), in order to obtain the polarisation fractions in the helicity 
frame we use the following expressions (for $W^+$): 
\begin{eqnarray}
f_0&=&2-5\langle \rm{cos}\theta^{*2}\rangle,\\
f_L&=&-\frac{1}{2}-\langle \rm{cos}\theta^*\rangle+\frac{5}{2}\langle \rm{cos}\theta^{*2}\rangle, \\
f_R&=&-\frac{1}{2}+\langle \rm{cos}\theta^*\rangle+\frac{5}{2}\langle \rm{cos}\theta^{*2}\rangle.
\end{eqnarray} 
These functions of $\theta^*$ can be used on an event-by-event basis as projections to extract the 
polarisation fractions for any $\sigma$ that does not depend on the kinematics of the individual leptons. 
In other words, the method assumes full acceptance for the leptons.

\begin{figure}[h]
\centering
\subfigure[]{
\includegraphics[trim=2.4cm 0 0 0,scale=0.56]{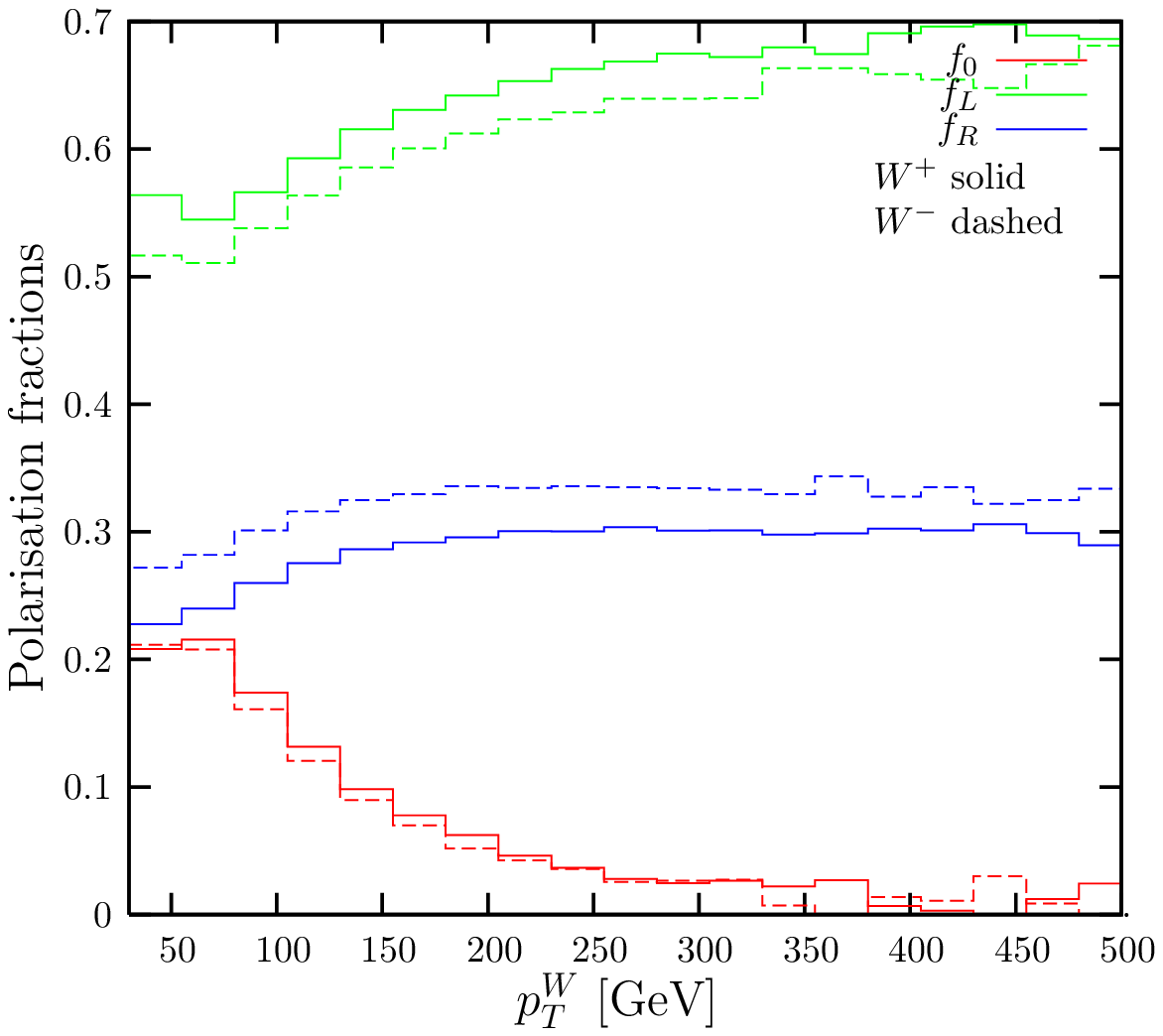}
}
\centering
\subfigure[]{
\includegraphics[trim=1.4cm 0 0 0,scale=0.56]{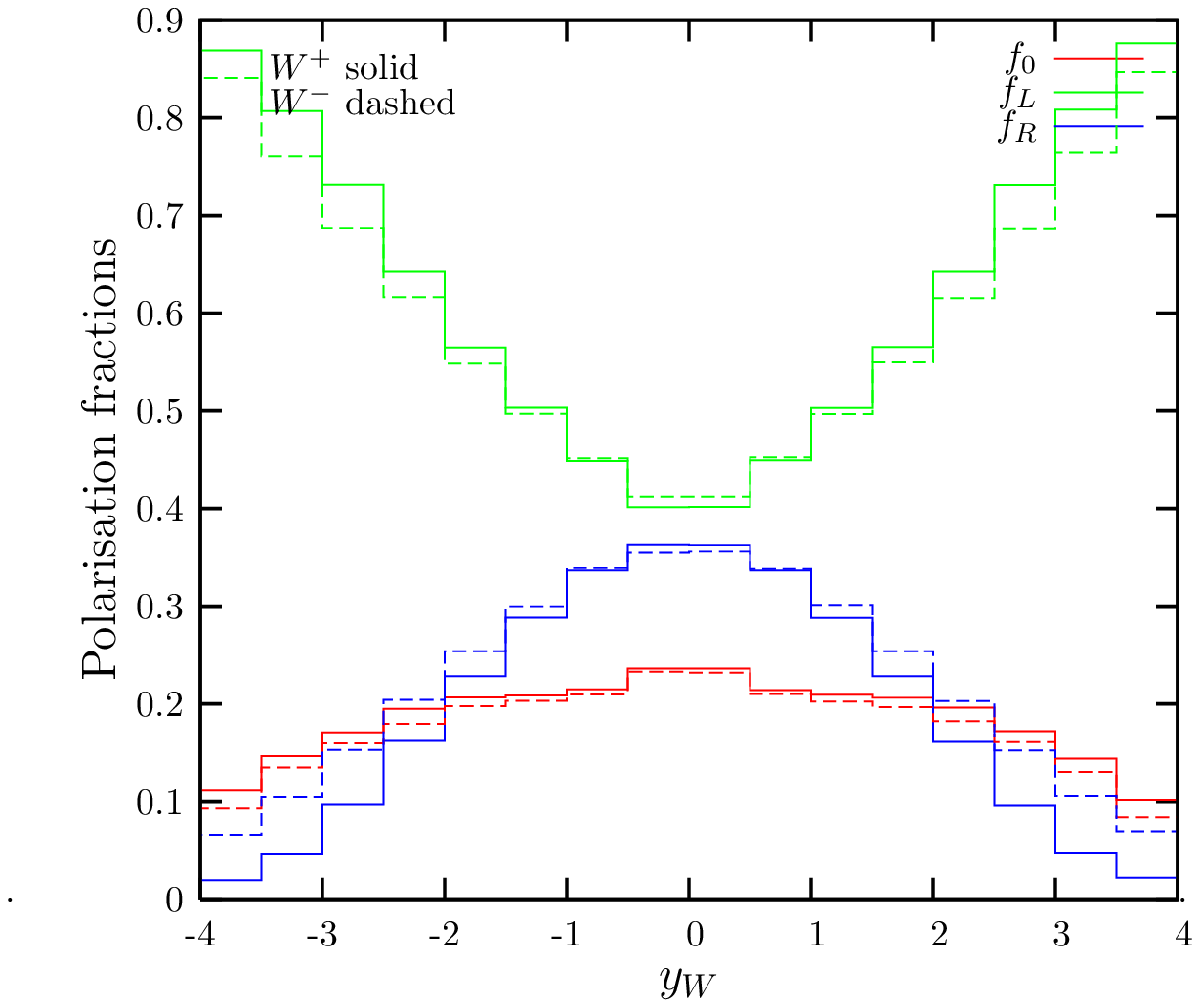}
}
\caption{Polarisation fractions as a function of a) $p_T^W$ and b) $y_W$ for 7 TeV and a jet
$p_T$ (=$p_T^W$ for $W+1$ jet) cut of 30 GeV for both $W^+$ and $W^-$.}
\label{fig7b}
\end{figure}

The polarisation fractions for $W$ bosons as a function of the $W$ boson transverse
momentum have already been obtained in \cite{Bern:2011ie} for $W+1$ jet with a simple jet
$p_T$ cut of 30~GeV. We reproduce these results (for the LHC at 7~TeV)
for both $W^+$ and $W^-$ in Fig.~\ref{fig7b} where we also show the results as a function of the $W$ rapidity. The small difference between the $W^+$ and $W^-$ polarisations
is due to the difference between the valence $u-$ and $d-$quark PDFs that
forces the $W^+$ to be slightly more left-handed. We also see the longitudinal fraction falling to zero at large transverse momentum, in agreement with the equivalence theorem~\cite{Cornwall:1974km}. 

\begin{figure}[h]
 \centering
 \includegraphics[trim=1.3cm 0 0 0,scale=0.6]{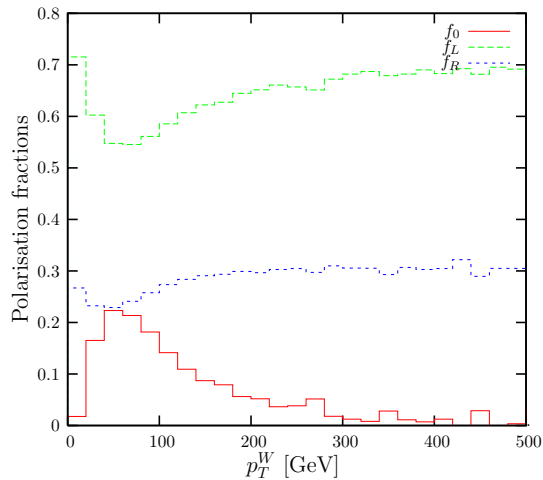}
 \caption{Polarisation fractions as a function of $p_T^W$ for 7 TeV and a nominal jet/parton 
$p_T$ cut of 0.1 GeV for $W^+$. }
 \label{lowpt}
 \end{figure}

In the case of $W+1$~jet production the cut on the jet transverse momentum effectively acts as a cut on
the $W$ $p_T$, which explains the cut-off at low $p_T^W$.
By decreasing the jet $p_T$ cut we obtain the limiting fractions for $p^W_T\sim 0$ 
shown in Fig.~\ref{lowpt}.\footnote{In practice we need to impose a small $p_T$ cut, as the cross section for 
$W+1$~jet production is formally divergent at zero $p_T$, even though the polarisation fractions are finite in this limit.} 
These are the same as for leading-order $W$ boson production along
the beam direction. In this case the polarisation of the $W$ is simply determined by
the momentum of the colliding quark and antiquark. The $W$ boson is left-handed if
the quark has more momentum than the antiquark, and right-handed otherwise.
Longitudinal polarisation is not permitted by angular momentum conservation. 
The exact values of the fractions at zero transverse momentum therefore depend on the relative values
of the corresponding quark and antiquark PDFs. At the LHC, quarks have on average more momentum than antiquarks
which explains the difference in values $f_L=0.73$ and $f_R=0.27$ for $W^+$, while the same arguments 
apply for $W^-$ leading to $f_L=0.68$ and $f_R=0.32$. At the Tevatron, quarks and
antiquarks have on average the same momentum and therefore we expect the
fractions to be closer to $0.5$ for $W^+$. The values are indeed calculated to be $f_L=0.60$ and $f_R=0.40$. 
These are  not exactly $0.5$, as the $W^+$ is primarily produced from a proton $u-$quark and an antiproton $\bar{d}$, 
with the $u-$quark carrying more momentum on average.  In this case the fractions for $W^-$ are exactly reversed, 
with $W^-$ being preferentially right-handed.

For $W+1$~jet production, different subprocesses contribute to the total cross section.
The analysis for $W+1$~jet production in \cite{Bern:2011ie} explains why given the dominant parton subprocesses
one expects $W$ bosons to remain predominantly left-handed at high transverse momentum. 
The argument is based on the spin$-1$ nature of $W$ and its coupling to left-handed fermions 
at the helicity amplitude level.

\begin{figure}[h]
\begin{minipage}[b]{0.5\linewidth}
 \centering
 \includegraphics[trim=1.3cm 0 0 0,scale=0.6]{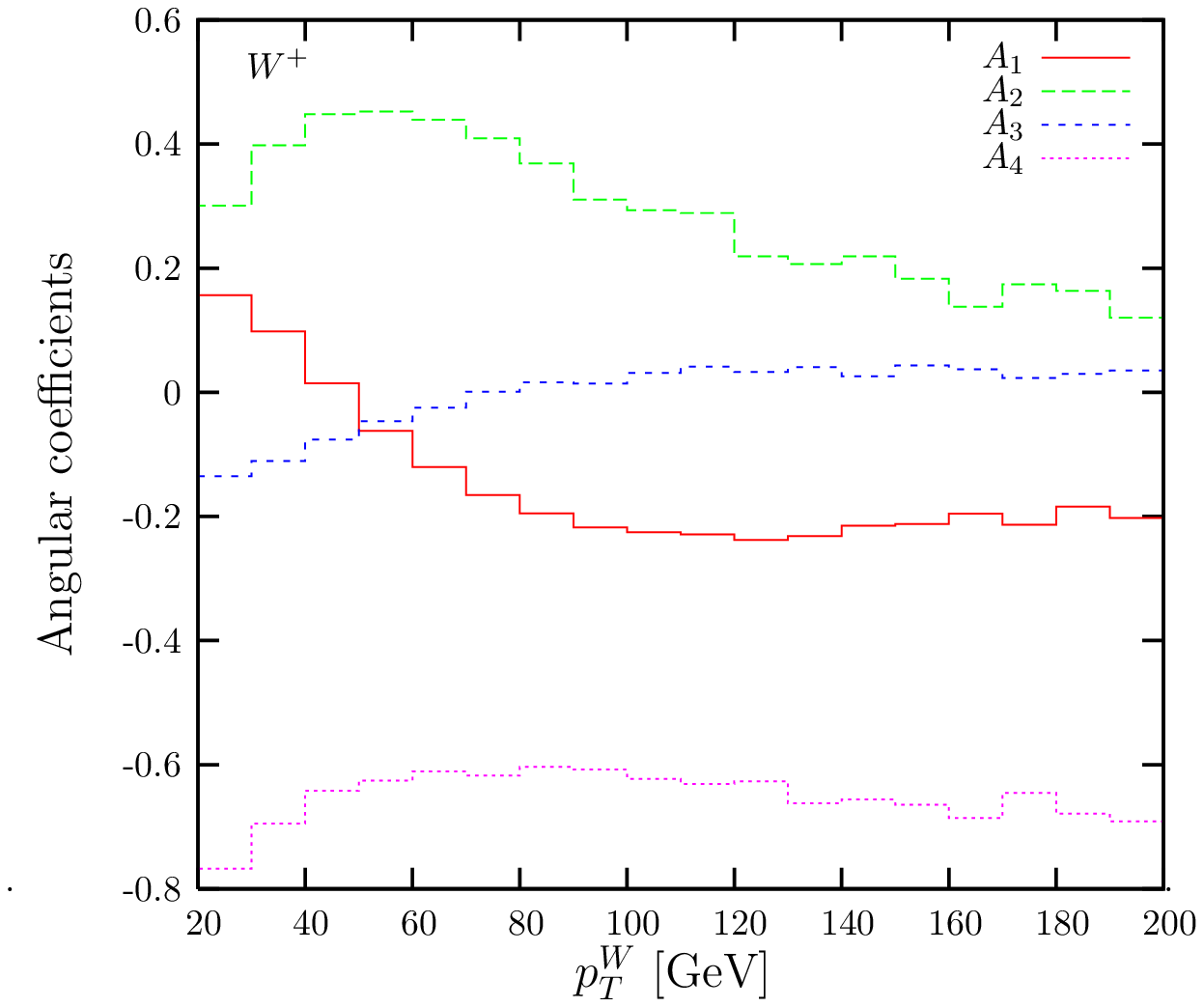}
\end{minipage}
 \hspace{0.5cm}
 \begin{minipage}[b]{0.5\linewidth}
\centering
 \includegraphics[trim=1.3cm 0 0 0,scale=0.6]{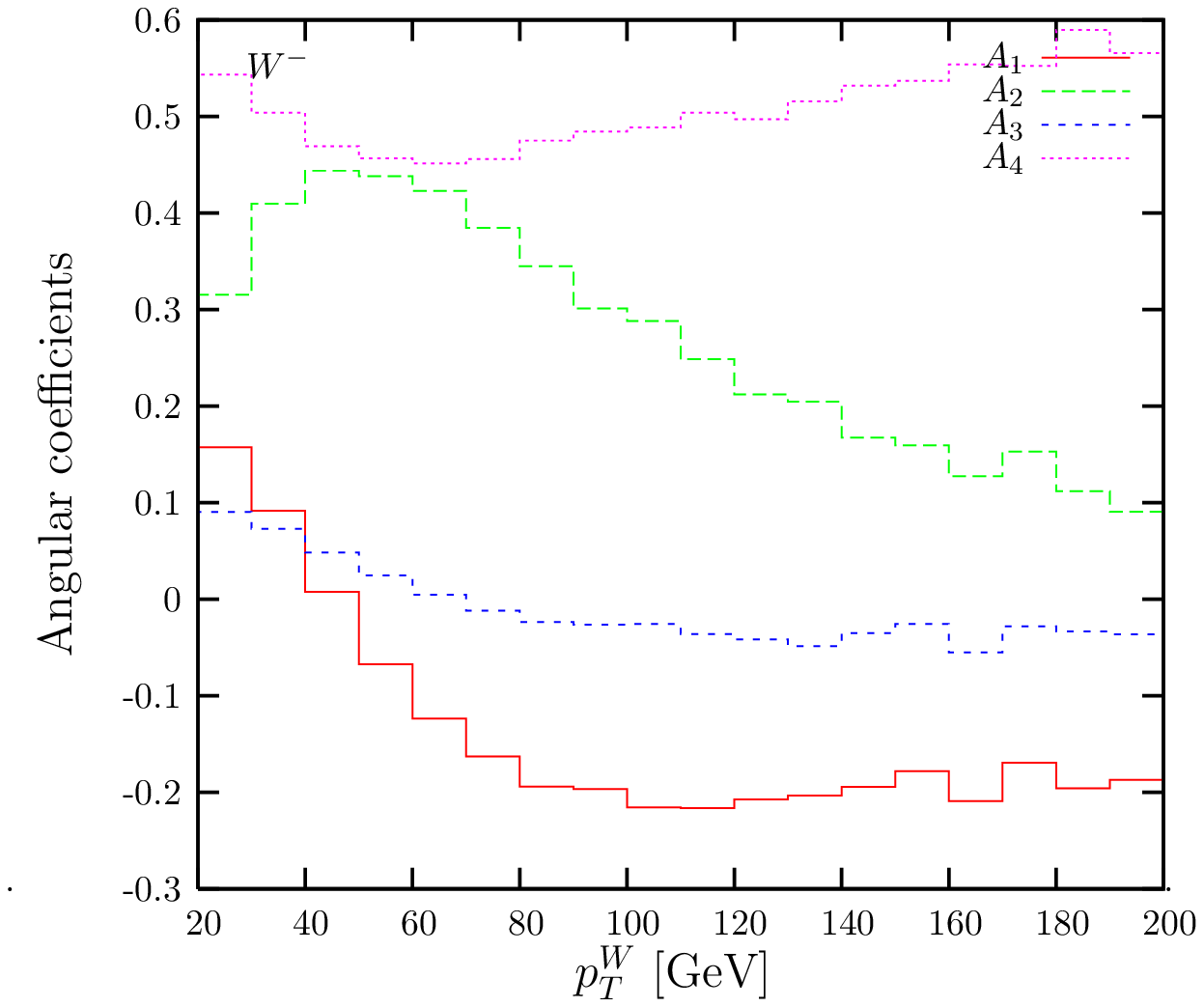}
\end{minipage}
 \caption{Angular coefficients  for $W^-$ and $W^+$ plus one jet with no imposed cuts 
in proton-proton collisions at 7 TeV. }
 \label{ang}
 \end{figure}

In addition to the polarisation fractions, which are the diagonal elements of the $W$ boson spin-density matrix, we can also
compute the full set of (polar and azimuthal) angular coefficients as a function of the $W$ transverse momentum 
as in \cite{Bern:2011ie} for the LHC and \cite{Mirkes:1994eb} for the Tevatron. The differential cross section is written as: 
\begin{eqnarray}\nonumber
\frac{1}{\sigma}\frac{d\sigma}{d\text{cos}\theta^*d\phi^*}&=&\frac{3}{16\pi}[(1+\text{cos}^2\theta^*)+A_0\frac{1}{2}(1-3\text{cos}^2\theta^*)+A_1 \text{sin} 2\theta^* \text{cos}\phi^*\\
&+&A_2\frac{1}{2}\text{sin}^2\theta^*\text{cos}2\phi^*+A_3\text{sin}\theta^*\text{cos}\phi^*+A_4\text{cos}\theta^*],
\label{fulldist}
\end{eqnarray} with the angle $\phi^*$ defined as in \cite{Bern:2011ie}. 
Integrating over $\phi^*$ from 0 to 2$\pi$ we recover Eq.~(\ref{diffeq}) and we can relate $f_{0,R,L}$ to 
the angular coefficients $A_i$. Three more coefficients not shown in Eq.~(\ref{fulldist}) vanish at LO because of parity invariance. 
The remainder of the coefficients can be determined using appropriate projections as for the polarisation fractions. 
We show the LO results obtained for $W+1$~jet for $W^-$ and $W^+$ in Fig.~\ref{ang}. At LO $A_0=A_2$~\cite{Lam:1980uc} 
and therefore we only show $A_2$ in the plots. The results are in good agreement with the NLO results of 
\cite{Bern:2011ie} which also include events with more jets. 

The above results were obtained using our own $W+$~jets LO programme. We have checked that our
results agree with a calculation using MCFM \cite{MCFM}, after matching the electroweak parameters, PDFs and scale choice.
 
\subsection{Effect of cuts on the polarisation results}
For a more realistic analysis one needs to introduce additional acceptance cuts on the final-state lepton transverse 
momenta and rapidities. Indeed such cuts will also serve
to reduce backgrounds in searches for New Physics. The impact of 
cuts has been first noted in \cite{Mirkes:1994eb}, where the angular
distribution of the weak boson decay products was studied including the effect of introducing 
kinematic cuts and taking into account the detector energy resolution. In this section we investigate the effect
of selection cuts on the angular distributions. Cuts are introduced in turn to
disentangle the effect of different cuts. The cuts introduced are: $p_T^j>20$~GeV,
 $p_T^{\ell}>20$~GeV, $E_T^m>20$~GeV and $|\eta_{\ell,j}|<2.5$. The rapidity cut is imposed on both leptons and jets. 
These are designed to mimick a `typical' set of cuts employed by the experimental collaborations to select $W$ boson events. 
The first distribution
 we consider is the charged lepton angular distribution $d\sigma/d\cos\theta^*$.
\begin{figure}[h]
\begin{minipage}[b]{0.5\linewidth}
 \centering
 \includegraphics[trim=1.3cm 0 0 0,scale=0.6]{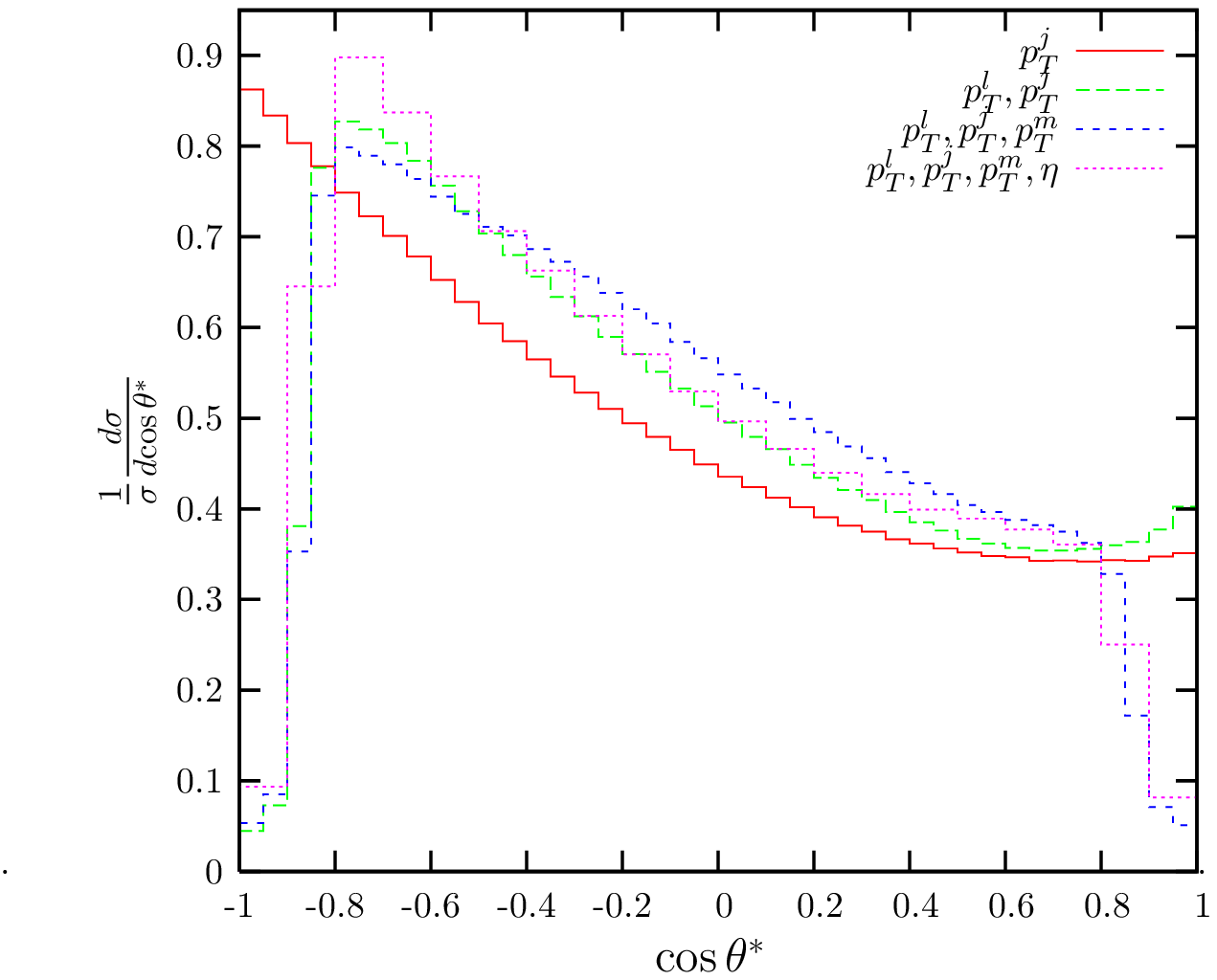}
 \caption{Normalised angular distribution for a set of different selection cuts imposed on final-state 
leptons and jets for $W^++1$~jet production at 7 TeV. }
 \label{coscuts}
\end{minipage}
 \hspace{0.5cm}
 \begin{minipage}[b]{0.5\linewidth}
\centering
 \includegraphics[trim=1.3cm 0 0 0,scale=0.6]{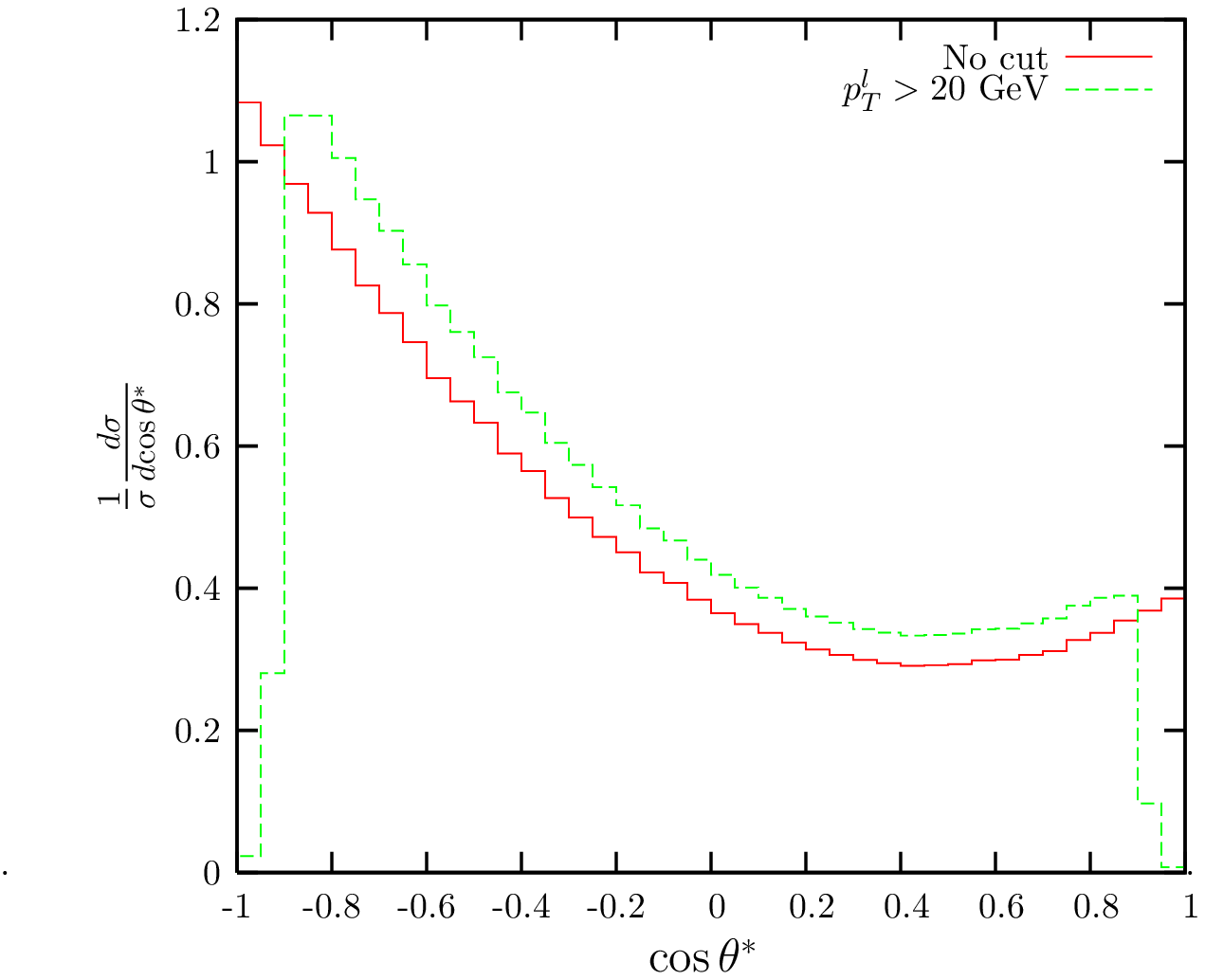}
 \caption{Angular distribution for LO $W^+$ production along the beam direction with no imposed cuts and with 
a cut on the charged lepton $p_T$. }
 \label{cos0j}
 \end{minipage}

 \end{figure}
The comparison between the normalised angular distributions for $W^+ + 1$~jet for different selection 
cuts is shown in Fig.~\ref{coscuts}. The cross section gets progressively smaller on the introduction 
of additional cuts but as we are considering the normalised distributions we only observe the changes in shape.

We note that for $W^++1$~jet production introducing a cut on the lepton transverse momentum reduces the 
cross section mainly in the region of $\theta^*\sim \pi$ and
 similarly the effect of a cut on the missing transverse energy is more important in the $\theta^* \sim 0$ region, 
as for a left-handed $W^+$ the neutrino is preferentially emitted in the direction of the $W$. For LO $W^+$ production along 
the beam direction with no associated jet the same effect is seen in the distribution as shown in Fig.~\ref{cos0j}. 
However in this case
 the transverse lepton momentum is exactly balanced by the missing transverse energy. 
Therefore the cut on the lepton transverse 
momentum implies the same cut on the missing transverse energy and the normalised angular distribution is modified at both ends. 
The $W$ rest frame is in this case identified as the centre-of-mass frame and it is therefore clear that imposing 
a cut on $p_T^{\ell}$ suppresses forward or backward scattering.

\begin{figure}[h]
 \centering
 \includegraphics[trim=1.3cm 0 0 0,scale=0.6]{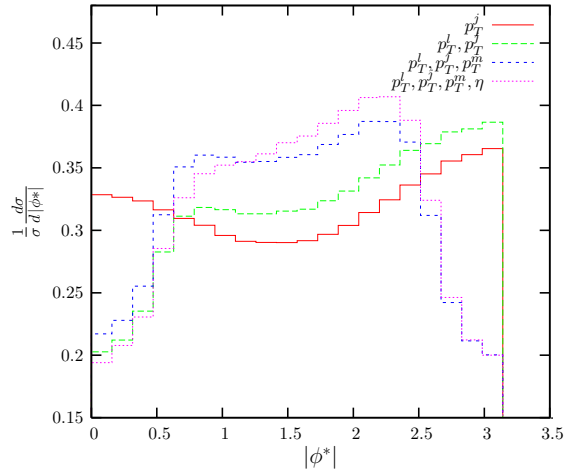}
 \caption{ Normalised azimuthal angle distributions for a set of different selection cuts 
imposed on final-state leptons and jets for $W^++1$~jet production at 7 TeV. }
 \label{phicut}
 \end{figure}

 One can also study the differential distribution in the azimuthal angle $\phi^*$ as defined in \cite{Bern:2011ie}. 
The distribution is even in $\phi^*$ and therefore the distributions shown in Fig.~\ref{phicut} correspond to 
$d\sigma/d|\phi^*|$.  Again we see that the acceptance cuts do not modify the cross section uniformly over the 
range of the angle $\phi^*$ and accordingly change the shape of the normalised distribution. 

\begin{figure}[h]
\centering
\subfigure[]{
\includegraphics[trim=1cm 0 0 0 ,scale=0.54]{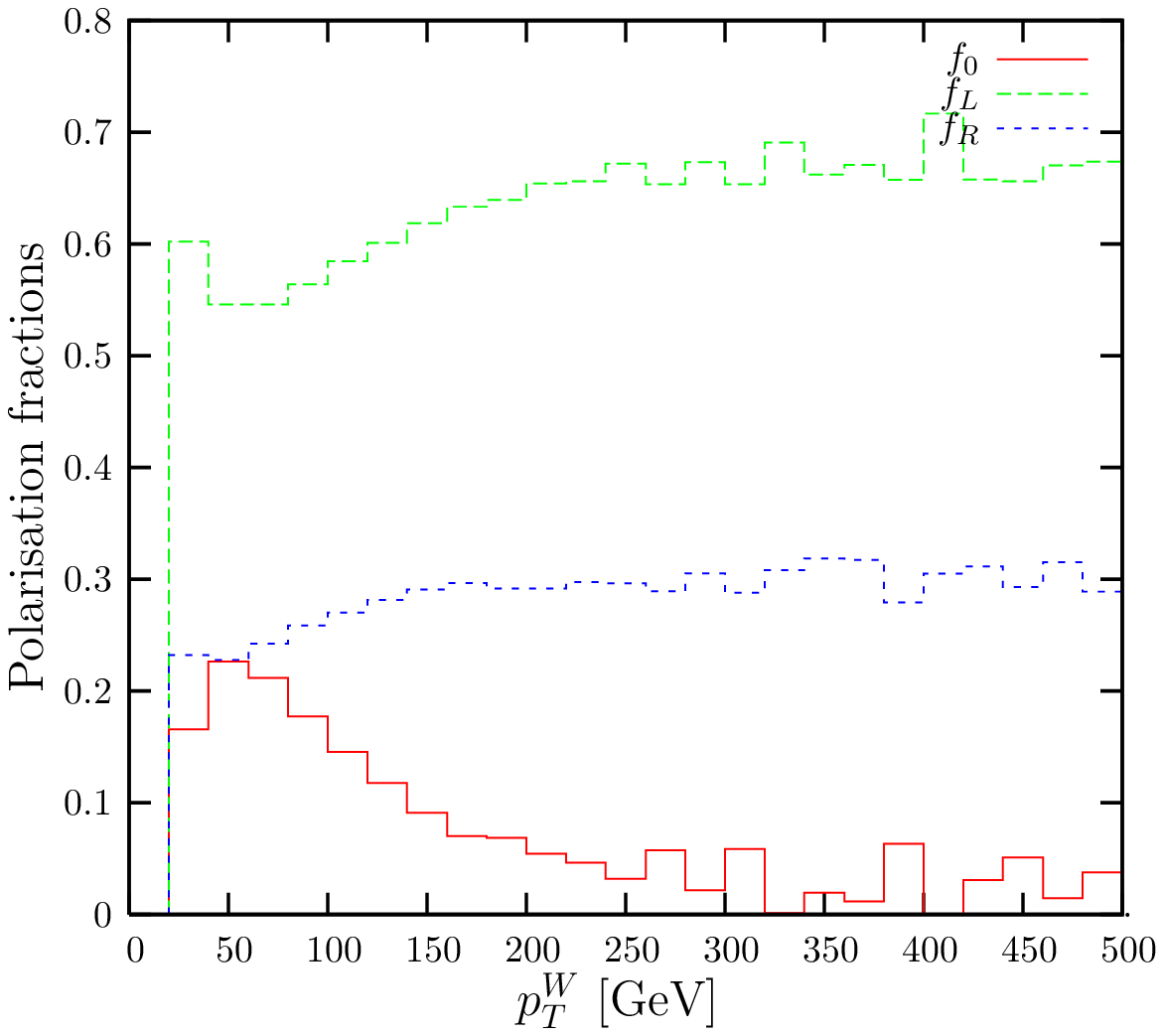}
}
\subfigure[]{
\includegraphics[trim=1cm 0 0 0 ,scale=0.54]{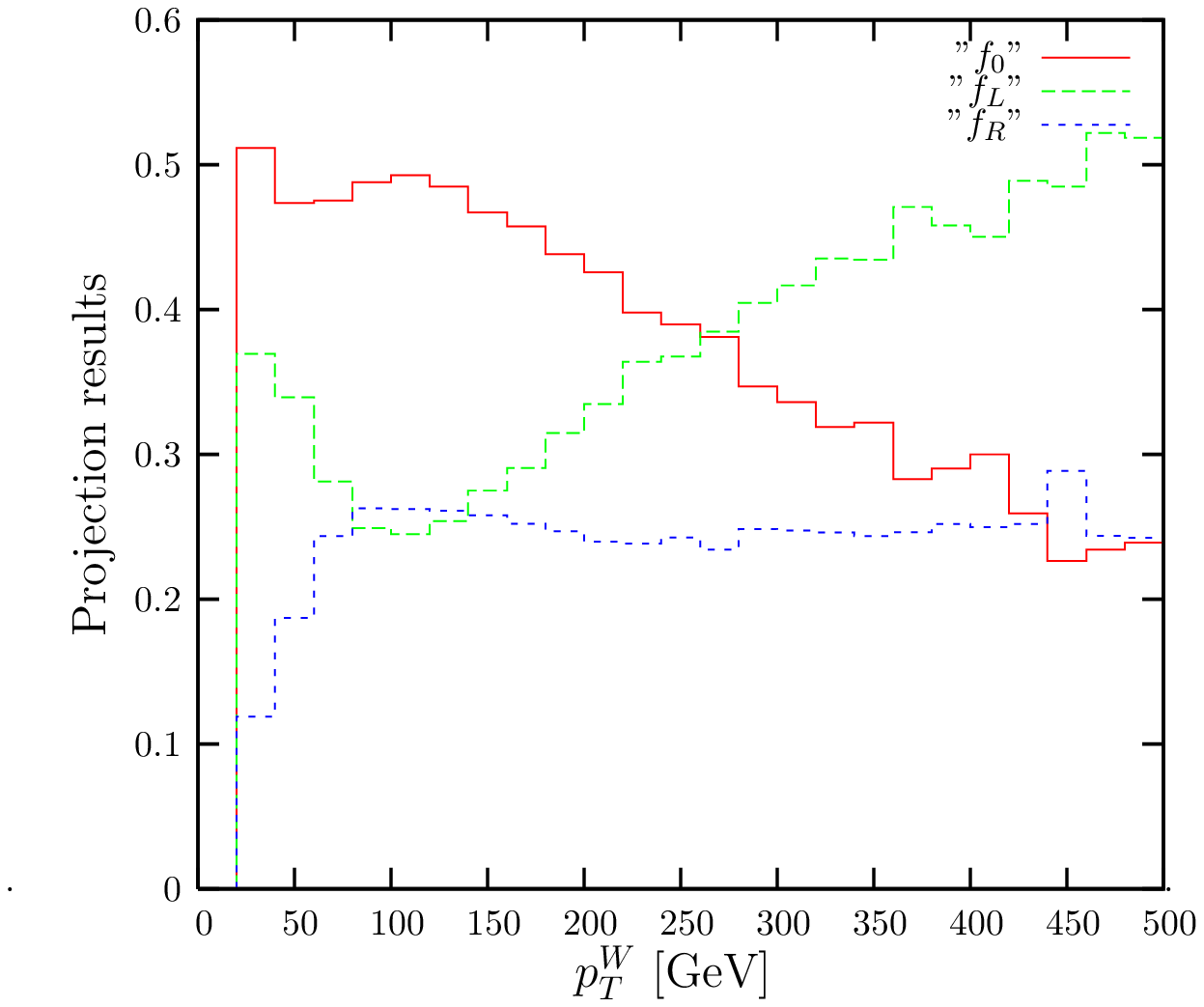}
}
\caption{Projection results as a function of $p_T^W$ for 7 TeV for $W^+$+1 jet with a jet
$p_T$ cut of 20 GeV and a) without and b) with a cut for charged
leptons $p_T^l>20$ GeV.}
\label{lepton} 
 \end{figure}

In general, we see that the angular distributions change rapidly
 on the introduction of additional cuts. Regarding the polarisation fractions, we note that no straightforward expression 
like that in Eq.~(\ref{diffeq}) applies once lepton cuts have been introduced, i.e. 
Eq.~(\ref{diffeq}) is valid only for full $(4\pi)$ lepton acceptance. One could still use the expressions for the fractions 
as observables, but these will no longer represent the polarisation fractions. 
Based on Fig.~\ref{coscuts}, if we insist on using the same $\theta^*$-dependent
 projections we expect them to give significantly different results. 
As an example, we show in Fig.~\ref{lepton} the result obtained for  a lepton transverse momentum cut of 20~GeV,
 in addition to the jet transverse momentum cut of 20~GeV, for $W^++1$~jet production at 7 TeV.

\begin{figure}[h]
 \centering
 \includegraphics[trim=1.3cm 0 0 0,scale=0.6]{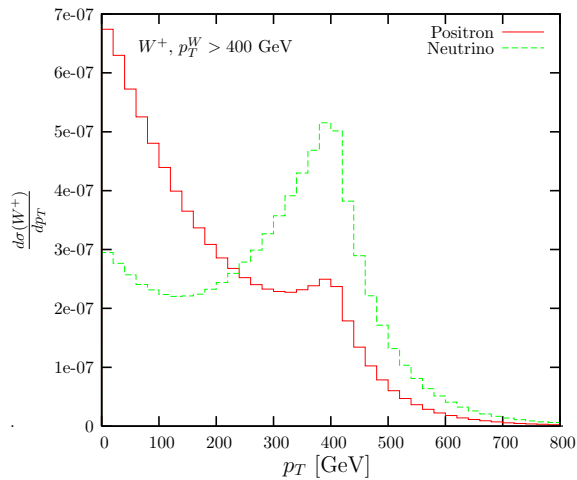}
 \caption{Lepton $p_T$ and missing transverse energy distributions for a high $p_T^W$ cut for $W^+ + 1$~jet production. }
 \label{ptdist}
 \end{figure}

In Fig.~\ref{lepton} we note the rather dramatic impact of the lepton cuts on the results.
Contrary to what one might naively expect, even a modest
lepton $p_T$ cut of 20~GeV modifies the results for $p^W_T$ up to about
500~GeV. This can be explained by the fact that $W$ bosons are predominantly left-handed at high $p_T^W$, 
creating an asymmetry in the distributions of lepton transverse momentum and missing transverse energy. 
The distributions of lepton transverse momentum and missing transverse energy are shown in Fig.~\ref{ptdist}
 for a cut of 400~GeV on $p_T^W$. Since the cross section is a rapidly falling function of $p_T^W$ most 
of the contribution comes from $W$ bosons with $p_T$ just above 400~GeV. 
The two distributions are peaked at opposite ends, with the neutrino
 along the $W^+$ direction most of the time as the $W^+$ is predominantly left-handed. For predominantly right-handed $W^+$ 
we expect  the peaks to switch positions and for a purely longitudinal $W^+$ we expect the same shape for both and a peak at
 around half the $p_T^W$ cut value. Similar arguments apply for $W^-$, but in this case the distributions are interchanged.
 We note that the shape of the lepton $p_T$ resembles the angular distribution 
$d\sigma/d{\cos}\theta^*$. This explains why the effect of a modest lepton $p_T$ cut extends to the region of very high $p_T^W$.

As already discussed above, the definition of the polarisation fractions was based on full acceptance in 
the azimuthal angle $\phi^*$ and in general no constraint on the kinematics of the individual leptons. 
Introducing lepton cuts leads to a dependence of the fractions on other components 
of the $W$ spin density matrix. 
Experimentally, lepton cuts are imposed to accommodate finite detector acceptance. 
In an analysis aimed at extracting the $W$ polarisation, experiments must correct for their 
lepton cuts before using the projections to obtain the polarisation fractions.

\begin{figure}[h]
\begin{minipage}[b]{0.5\linewidth}
 \centering
 \includegraphics[trim=2.4cm 0 0 0,scale=0.55]{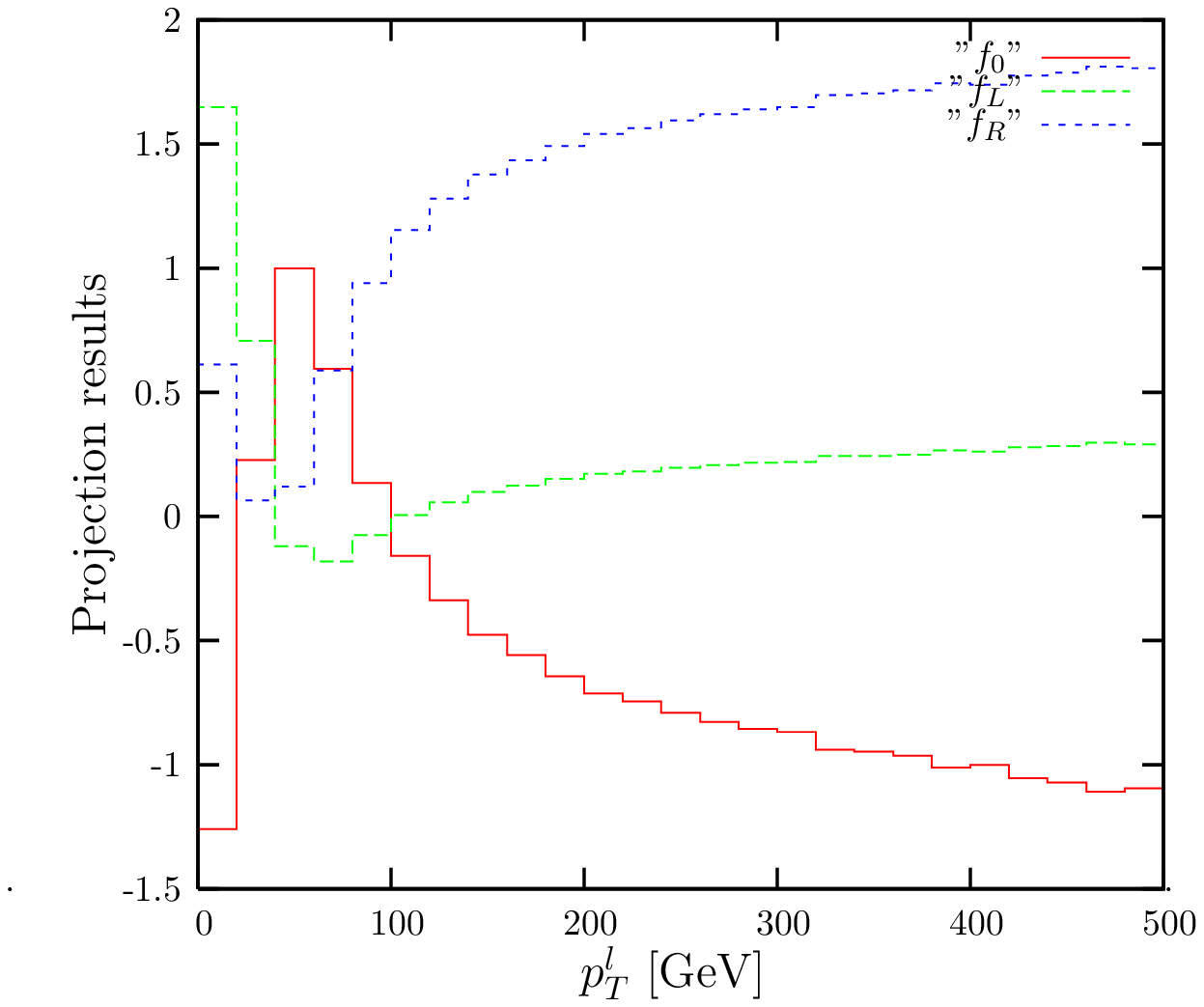}  
\end{minipage}
 \hspace{0.5cm}
 \begin{minipage}[b]{0.5\linewidth}
 \includegraphics[trim=2.4cm 0 0 0,scale=0.55]{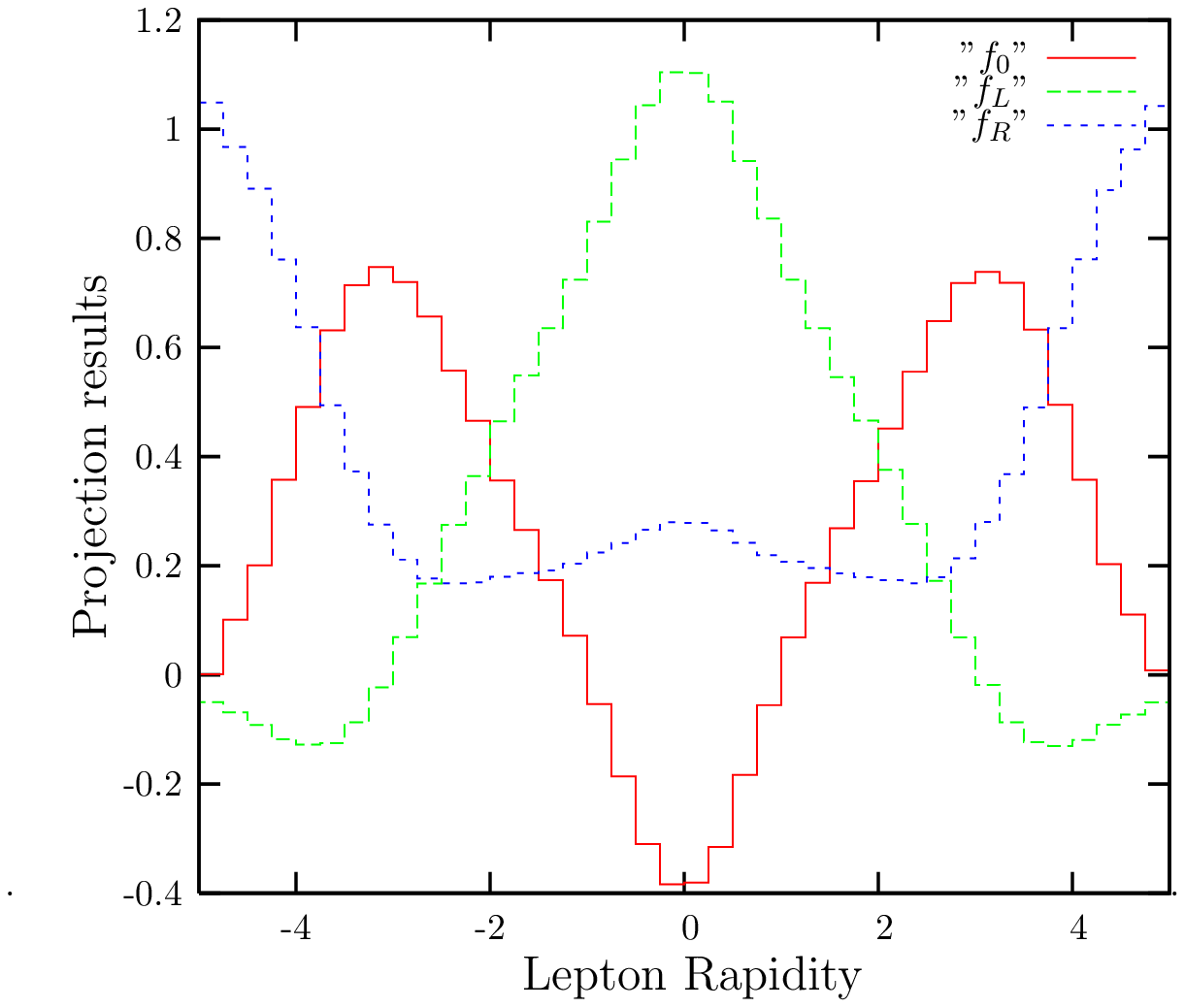}
\end{minipage}
\caption{Projection results for $W^+$ as a function of the lepton $p_T$ and
rapidity with $p^j_T>20$~GeV.}
 \label{leppt}
\end{figure}

In principle we could still use the expressions for the projections as observables even when lepton cuts have been applied.
In this case in order to explain the impact of cuts we need to study the dependence on the observables 
on which we impose the cuts, e.g. the
lepton rapidity and the lepton transverse momentum. The results for the lepton
transverse momentum and rapidity are shown in Fig.~\ref{leppt}. 
We see in these plots that ``$f_0$'',``$f_L$'' and ``$f_R$'' are rapidly
changing functions of the two observables which is reflected in the severe
impact of the cuts on the results. 
We note here that once we consider the projections as a function of other variables which depend 
on the individual kinematics of the leptons we observe that these can become negative. 
This is not surprising, as these expressions no longer represent the polarisation fractions.

The results for the expressions (not the polarisation fractions except for rows 1 and 2)
for different cuts are given in Table~\ref{fractions}. 

\begin{table} 
\begin{center}
    \begin{tabular}{ | c | c | c | c |}
    \hline
   Cuts  & ``$f_0$'' & ``$f_L$'' & ``$f_R$''   \\ \hline
    $p_T^j>$ 30~GeV & 0.20 & 0.56 & 0.23 \\ \hline
     $p_T^j>$ 20~GeV & 0.18 & 0.59 & 0.23 \\ \hline
$p_T^j>$ 20~GeV, $p_T^l>$ 20~GeV & 0.50 & 0.35 & 0.15 \\ \hline
    $p_T^j>$ 20~GeV, $p_T^l>$ 20~GeV, $p_T^m>$ 20~GeV & 0.68 & 0.29 & 0.03 \\ \hline
$p_T^j>$ 20~GeV, $p_T^l>$ 20~GeV, $p_T^m>$ 20~GeV, $|\eta_{l,j}|<$2.5 & 0.59 & 0.36 & 0.05 \\ \hline
    
   \end{tabular}
\end{center}
\caption{Comparison of results for the different cuts. The values represent polarisation fractions only for the first two rows.}
\label{fractions}
\end{table}

\begin{figure}[h]
\centering
\subfigure[]{
\labellist
\footnotesize\hair 1.5pt
\pinlabel {$p_T^W>50$ GeV} at 140 646
\endlabellist
 \centering
 \includegraphics[scale=0.56]{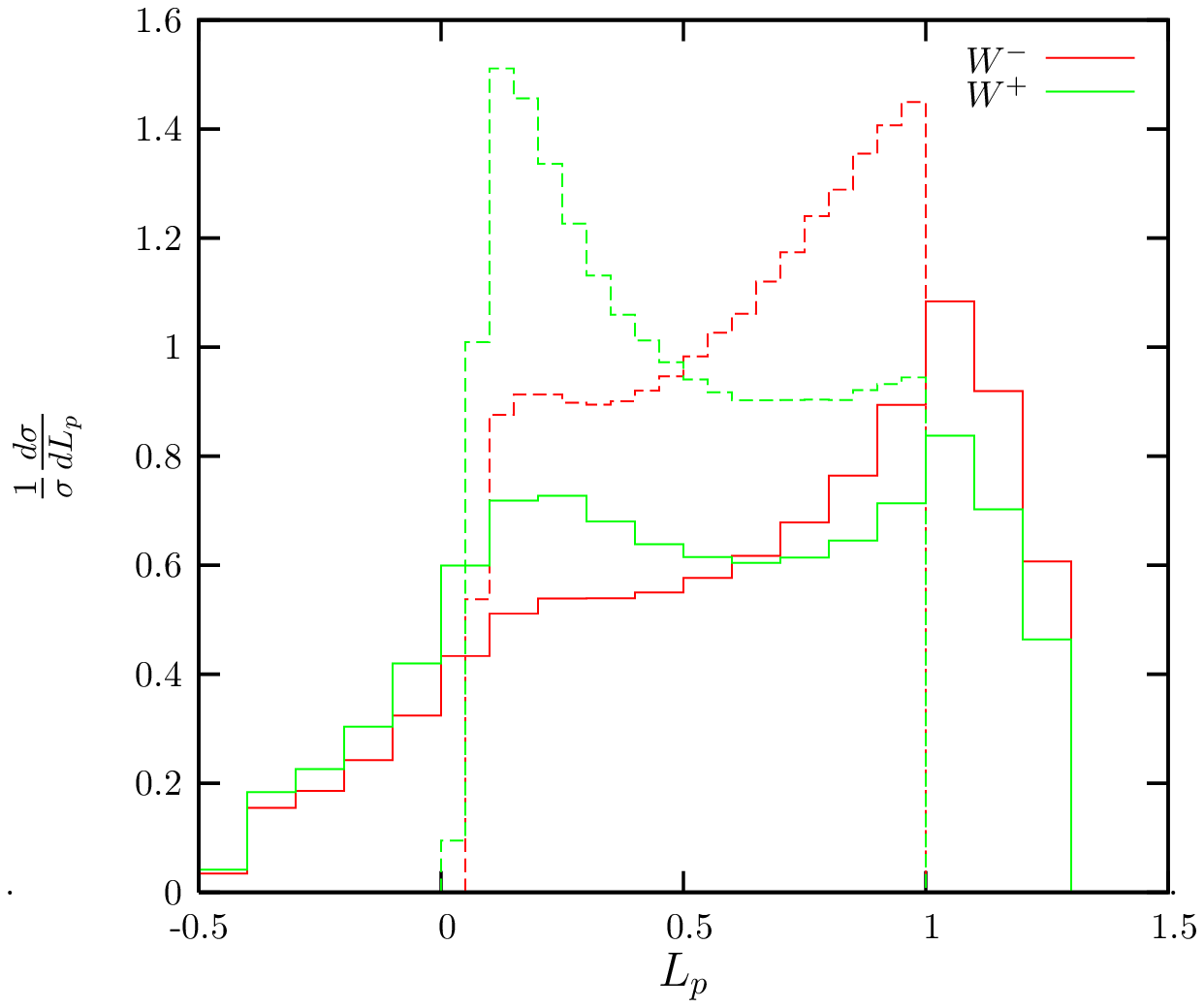}
}
\subfigure[]{
\labellist
\footnotesize\hair 1.5pt
\pinlabel {$p_T^W>100$ GeV} at 140 646
\endlabellist
 \includegraphics[scale=0.56]{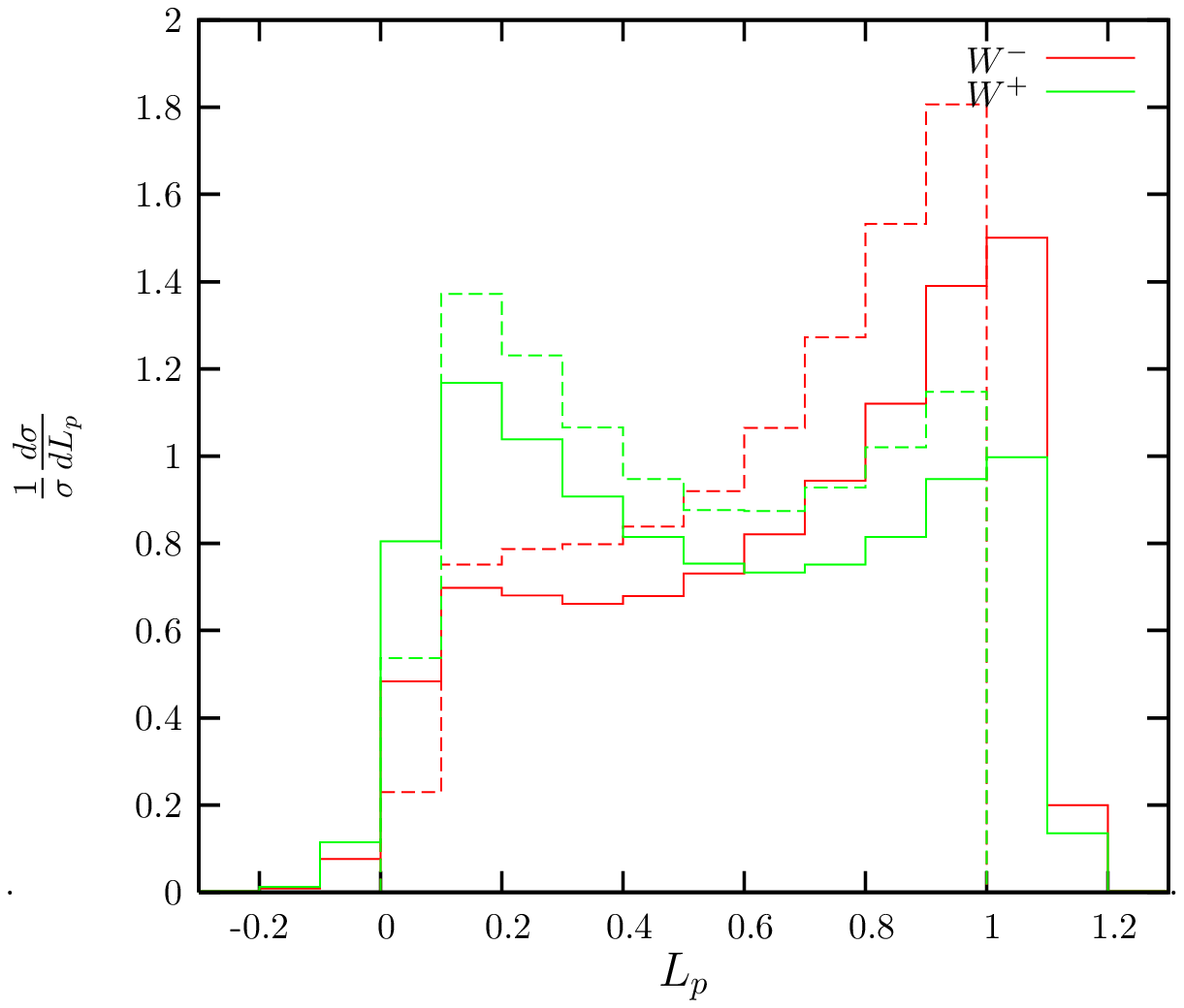}
}
 \caption{Normalised distributions for the variable $L_p$ for $W^-$ and $W^+$ with the CMS cuts. $P^T_W$ cut increased from a) 50 to b) 100 GeV. Solid lines show $L_p$ as defined above while dashed lines show $L_p^h$.}
\label{Lp}
 \end{figure}

The above predictions are obtained using a Monte Carlo event generator programme from which 
we know exactly the momentum of the $W$ boson. 
 Of course  experimentally it is impossible to reconstruct exactly and unambiguously the momentum of the $W$ boson, as one can 
only measure the transverse momentum of the neutrino, unless a $W$ mass contraint is applied on the neutrino-electron 
pair (see for example the Tevatron
 polarisation studies in top pair production in \cite{Abazov:2010jn}). Even when the extra mass constraints are applied, 
an ambiguity remains and a further selection needs to be made.
  Therefore a straightforward extraction and use of the angle 
$\theta^*$ to define the polarisation fractions is not feasible. The first measurement of the
polarisation of the $W$ boson by CMS~\cite{Chatrchyan:2011ig}
 introduced the variable $L_p$, defined as: 
\begin{equation}
L_p=\frac{\vec{p}_T(l)\cdot \vec{p}_T(W)}{|\vec{p}_T(W)|^2},
\end{equation} 
where all quantities can in principle be reconstructed. In the limit of very high $W$ transverse momentum 
$\cos\theta^*=2(L_p-1/2)$. For purposes
 of comparison with CMS, we also calculate the distribution of $L_p$, applying the cuts given in \cite{Chatrchyan:2011ig} 
for the muon channel. The cut imposed on $p_T^W$ is 50~GeV. We note that to approach as closely as possible the CMS 
results we prefer the muon channel, as it is less affected by backgrounds (as shown in \cite{Chatrchyan:2011ig}).
 We note that these are parton-level results, with no detector simulation obtained for $W+1$~jet. 
The experimental results include events with more jets, as there is no specific jet number requirement. 
The normalised distributions for $W^-$ and $W^+$ are shown in Fig.~\ref{Lp}. 
 The shape of the $L_p$ distributions agrees reasonably 
well with the results of Fig.~2 in the CMS paper. In the plot we also show the distribution of 
$L_p^h=(\cos\theta^*+1)/2$ to investigate the possibility of using $L_p$ to recover $\cos\theta^*$.
 The CMS analysis imposes a cut of 50~GeV on the $W$ transverse momentum but the two distributions differ
 significantly, with the experimental distribution extending above one and below zero. 
Agreement improves for a $p_T^W$ cut of 100~GeV, as also shown in Fig.~\ref{Lp}~b). 
The results for the polarisation fractions in \cite{Chatrchyan:2011ig} are obtained using the template 
method to extract the distributions of $L_p$ for pure longitudinal, left- and right-handed samples and then fitted to the data to obtain the polarisation fractions. This accounts both for the effect of the selection cuts 
and the mismatch of $L_p$ and $L_p^h$, as  shown in Fig.~\ref{Lp}.

\subsection{$W$ plus more jets production}
The results of the previous section have been obtained by considering $W+1$~jet production. 
Similar considerations can be made for $W+n$~jets with $n\geq 2$. 
The $W+n$~jets cross section falls with increasing number of jets, as each additional jet is associated with
 an extra power of $\alpha_S$.  
At the 7~TeV LHC, with a cut of 30~GeV on the transverse momentum of all jets,
 we obtain the following LO results for $\sigma( W^+ + n$~jets$)$ for the electron decay channel 
(with no cuts on the leptons or missing energy): 0.611~nb for one jet, 0.215~nb for 2 jets and 0.0741~nb for 3 jets. 
These results are obtained by running separately the $W+1$, 2 and 3 jets routines and checking that all produced jets 
pass the $p_T$ cut. In practice the experimental $W+1$~jet sample will contain events in which two jets 
were emitted but one of them falls outside the acceptance region of the detector etc. 
Effective use of MC generators is needed to correctly account and correct for this effect.

\begin{figure}[h]
 \centering
 \includegraphics[trim=1.3cm 0 0 0,scale=0.6]{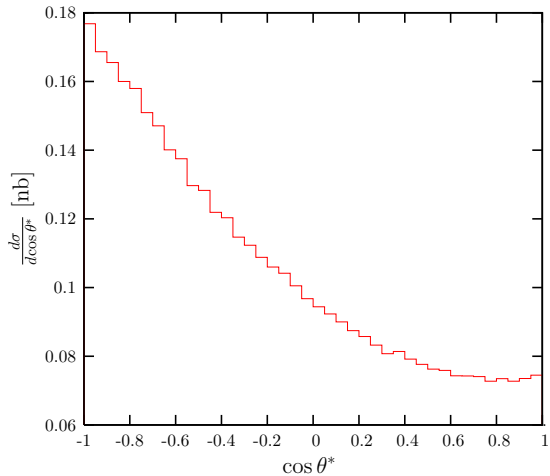}
 \caption{Angular distribution $d\sigma/d\cos\theta^*$ for $W^+ +2$~jets with $p_T^j>$30~GeV at 7~TeV.}
\label{angle}
\end{figure}

Similarly to the $W+1$~jet analysis, we study the lepton angular distribution in
the $W$ rest frame as shown in Fig.~\ref{angle} for $W^+ +2$~jets for a cut of 30~GeV on $p_T^j$. 
We note that the shape is identical to that for $W^++1$~jet. The polarisation fractions are calculated 
using the same method as for $W+1$~jet. An additional cut that needs to be set for two or more jets is a cut on the 
separation of two jets in order for them to be considered distinct jets. The cone separation variable $R$
has a negligible effect on the
polarisation fractions.  Moreover we have checked using MCFM that NLO corrections 
have no sizable impact on the polarisation fractions. The definition of the polarisation 
fractions as ratios over the total cross sections helps reduce the sensitivity to the NLO corrections. 
In \cite{Bern:2011ie} it has been shown using SHERPA~\cite{Gleisberg:2003xi} that the results remain stable 
even when parton shower effects are taken into account. 

\begin{figure}[h]
 \centering
 \includegraphics[trim=1.3cm 0 0 0,scale=0.6]{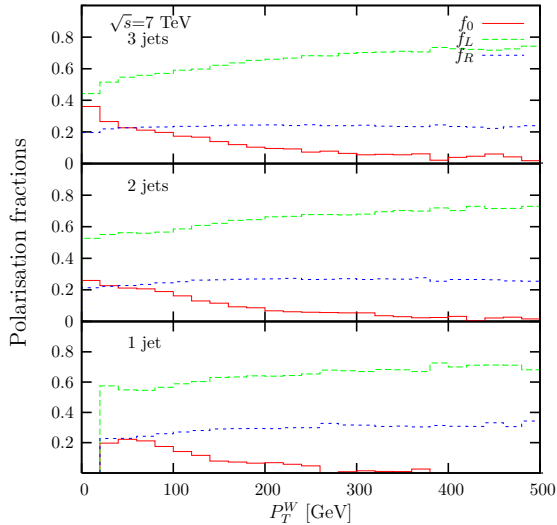}
 \caption{Polarisation fractions for $W^+$+1, 2 and 3~jets with $p_T^j>30$~GeV at 7~TeV obtained using MCFM.}
\label{morejets}
 \end{figure}

For comparison we collect in Fig.~\ref{morejets} the results of the polarisation fractions for 1, 2 and 3 jets with a jet transverse momentum 
cut of 30~GeV. As already noted in
\cite{Bern:2011ie}, even though the kinematics 
become more complicated with an increasing number of jets the polarisation fractions are not very sensitive to the number
of jets, with changes only observed at low $W$ transverse momentum and $W$ bosons remaining predominantly left-handed 
at large transverse momentum.

\section{$W$ bosons from top pair production}
In addition to the production of $W$ bosons in association with QCD jets,
 the polarisation properties of $W$ bosons from other sources can also be investigated. 
The polarisation of $W$ bosons from top pair production has been measured at the  Tevatron both by 
CDF \cite{Aaltonen:2008ei,Aaltonen:2010ha} and D0 \cite{Abazov:2007ve,Abazov:2010jn}. 
The projections used are defined in \cite{AguilarSaavedra:2006fy}, where the polarisation of $W$ bosons from top decays has 
been used as a probe of anomalous $Wtb$ couplings.  The projections differ from those used for $W+$~jets, 
as the angle $\theta^*$ is defined in the $W$ rest frame relative to the $W$ direction in the top rest frame.

We begin by employing the projections of \cite{AguilarSaavedra:2006fy} to
reproduce the overall SM polarisation fractions of $W$ bosons from top pair production
given in \cite{AguilarSaavedra:2006fy} and measured at the Tevatron. The overall results show that for $W^+$ bosons 
from top pair production $f_0=0.70$ and $f_L=0.30$. The results can also be extracted from analytic expressions 
involving the $W$ and top masses (for massless $b-$quarks):
\begin{equation}
f_0=\frac{m_t^2}{m^2_t+2m_W^2} \,\,\,\,\, \textrm{and} \,\,\,\, f_L=\frac{2m_W^2}{m^2_t+2m_W^2}.
\end{equation}

\begin{figure}[h]
 \centering
 \includegraphics[trim=2.4cm 0 0 0,scale=0.6]{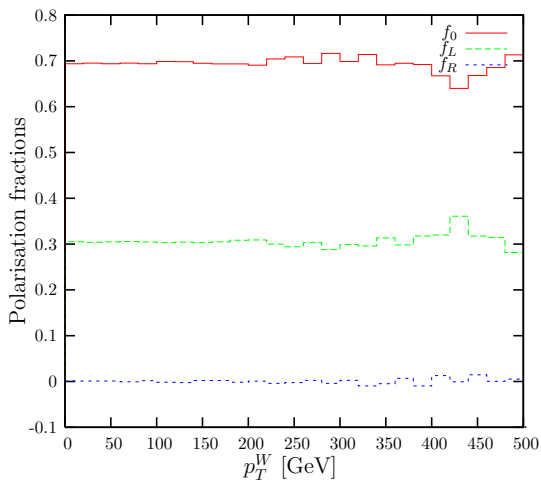}
 \caption{Polarisation fractions for $W^+$ from top decays with no imposed cuts. }
 \label{Wtop}
 \end{figure}

 The fractions obtained from the Monte Carlo simulation are shown as a function of the $W$ transverse momentum in Fig.~\ref{Wtop},
where no acceptance cuts have been imposed. For $W^-$ the result is $f_0=0.70$ and $f_R=0.30$. 
In contrast to $W+$~jets production we notice that the polarisation fractions are constant and there is essentially
 no dependence of the polarisation fractions on the $W$ $p_T$. 
This follows naturally from the way polarisation is defined using the top rest frame. 

\begin{figure}[h]
 \centering
 \includegraphics[trim=1.8cm 0 0 0,scale=0.6]{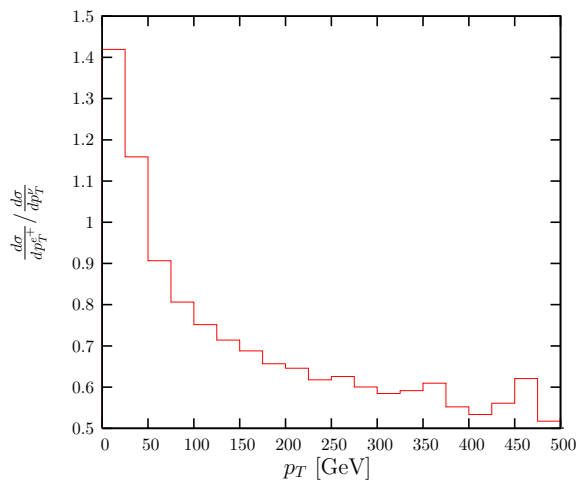}
 \caption{Ratio of differential cross sections for the
charged lepton  $p_T$ and the missing transverse energy for $W^+$ production from top decay at the LHC. 
The plot is identical for $W^-$. A jet cut $p_T^j>30$ GeV has been imposed.}
 \label{tdistr}
 \end{figure}

Unlike in  the case of $W+$~jets production,
in this process $W^+$ and $W^-$ are exactly equivalent as these are always produced in pairs from the decaying top--anti-top 
pair, hence the polarisation
fractions are related by $f_0^+=f_0^-$ and $f_R^+=f_L^-$. As for  $W+$~jets production we are interested in
identifying how the polarisation of the $W$ transforms into an asymmetry in the charged lepton and neutrino kinematics visible in the 
shape of the distributions. We show the ratio of
the $W^+$ lepton transverse momentum and missing energy distributions in
Fig.~\ref{tdistr}. Considering the symmetric production mechanism this ratio satisfies 
\begin{equation}
\frac{{d\sigma(t\bar{t})}/{dp_T^{e^+}}}{{d\sigma(t\bar{t})}/{dp_T^m}}
=\frac{{d\sigma(t\bar{t})}/{dp^{e^-}_T}}{{d\sigma(t\bar{t})}/{dp_T^m}}.
\end{equation}
The ratio of the charged lepton $p_T$ and
missing transverse energy distributions obtained for events where the $W^+$ decays leptonically and the 
$W^-$ hadronically is identical to that where the $W^-$ decays leptonically and the $W^+$ hadronically. 
Here we assume that only one of the two $W$ bosons decays leptonically so that there is only one charged lepton produced
in the event. We note that the shape of the ratio of the distributions is similar to that of $W^+$ in Fig.~\ref{lptmet}. 
This is expected as the asymmetry between the lepton transverse momentum and the missing transverse energy 
originates in the difference between the two transverse polarisation fractions. 
In both cases the produced $W^+$ is preferentially left-handed, which explains the decrease of the ratio with 
increasing $p_T$. In contrast, the produced $W^-$ is preferentially right-handed, leading to an identical ratio 
of distributions as for $W^+$ decays from top but fundamentally different from that of $W^- +$~jets production.
 
\begin{figure}[h]
 \centering
\subfigure[]{
 \includegraphics[trim=0.4cm 0 0 0,scale=0.62]{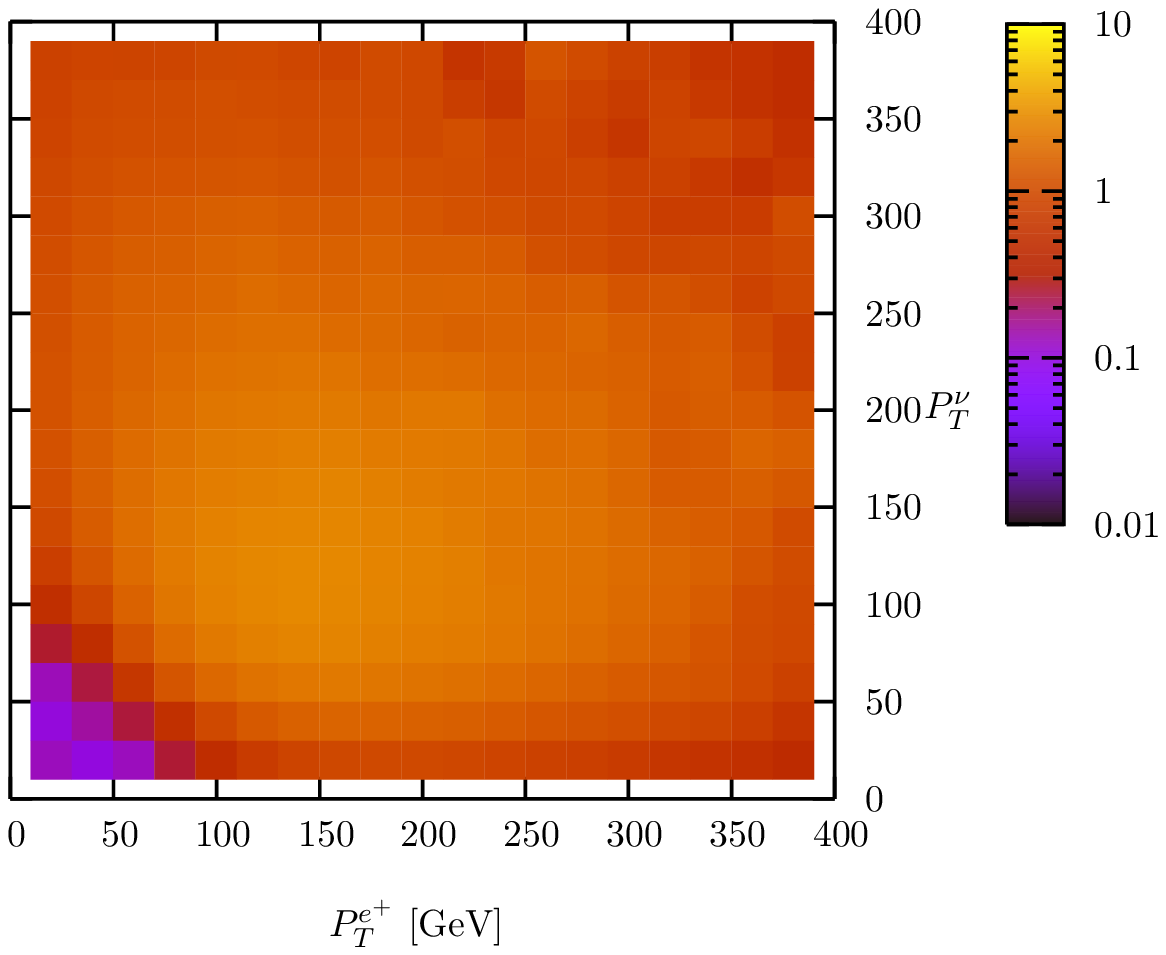}
 }
\subfigure[]{
 \includegraphics[scale=0.62]{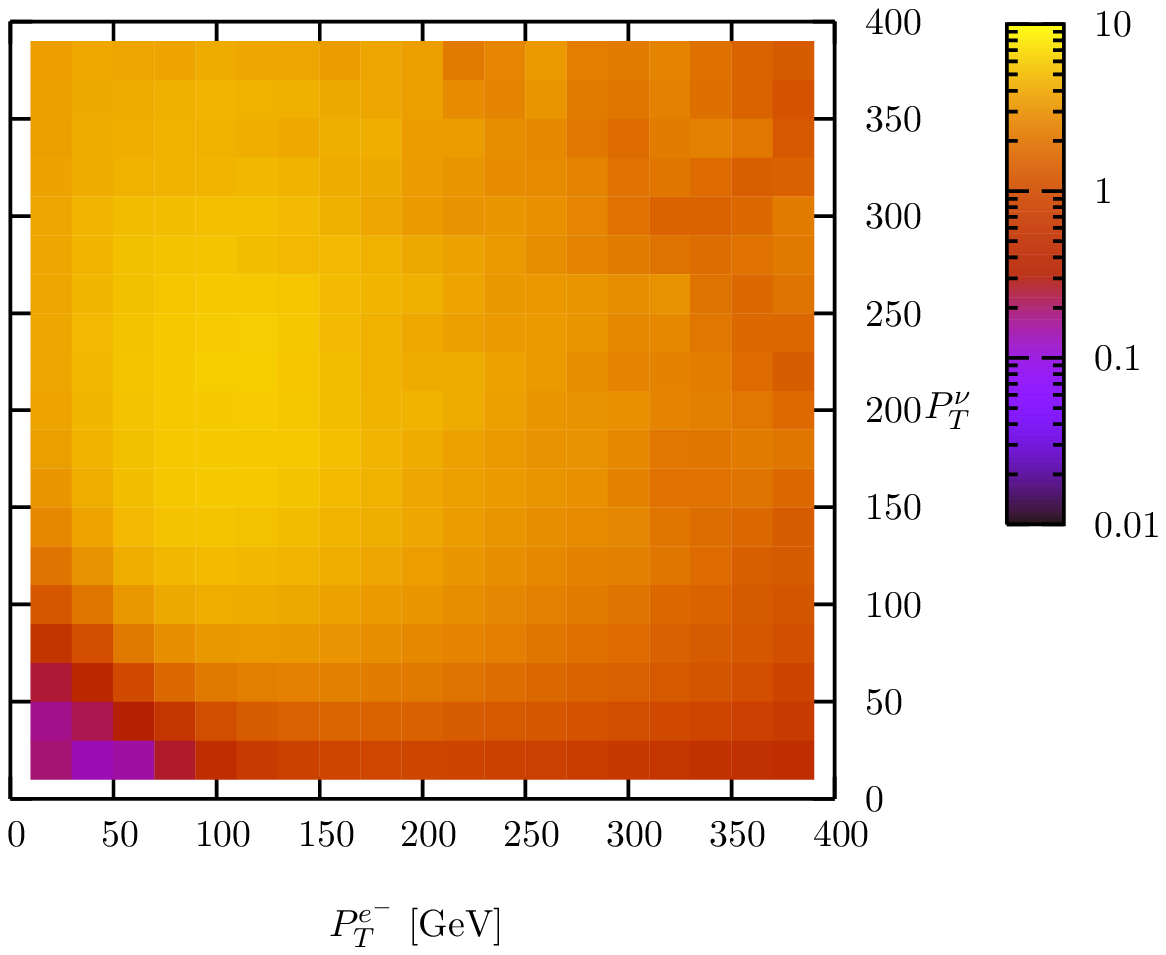}
 }
\caption{Density plots of the ratio of the double differential cross section for $t\bar{t}$ to $W+$~jets 
for a) $W^+$ and b) $W^-$. }
\label{dens}
 \end{figure}

We note that in this case no asymmetry is expected
between the number of produced electrons and positrons. Top pair production with one leptonic and one hadronic decay 
with no other mass selection cuts can be regarded as a background to QCD $W+4$~jets production (or {\it vice versa}). 
A comparison of the lepton and missing energy distributions in  $W+$~jets and $t\bar{t}$ production
has been made in \cite{Bern:2012te} where a density plot is used to show
 the difference between the two processes. The density plot is made using the $W+3$~jets 
and $W$ from top pair production with just three identified jets. Here we attempt to extract similar density plots using our 
$W+1$~jet and top pair production programs with no specific requirement on the number of jets. 
For $W+$~jets the characteristic shape is expected to persist when changing the number of jets 
as this is determined by the polarisation properties which are found to be stable. 
The ratio of the double differential cross section in lepton $p_T$ and missing transverse momentum 
($d^2\sigma/dp_T^ldp_T^m$) of $W$ from top decays to that of QCD $W+$~jets is shown in the density plots of Fig.~\ref{dens}. 
The same conclusions as in  \cite{Bern:2012te} can be drawn, even though we note some differences at low $p_T$ 
which are most likely related to the different number of jets in the process and the choice of selection cuts. 
In the density plots we note the asymmetry between $W^-$ and $W^+$ as the missing $E_T$ and $p_T^l$ space is 
populated differently by $W+$~jets and $t \to W$ events. We also note that the most noticeable difference occurs 
for  $W^-$ as the polarisation changes from left-handed in  $W+$ jets to right-handed in top pair production.

\begin{figure}[h]
 \centering
 \includegraphics[trim=1.3cm 0 0 0,scale=0.6]{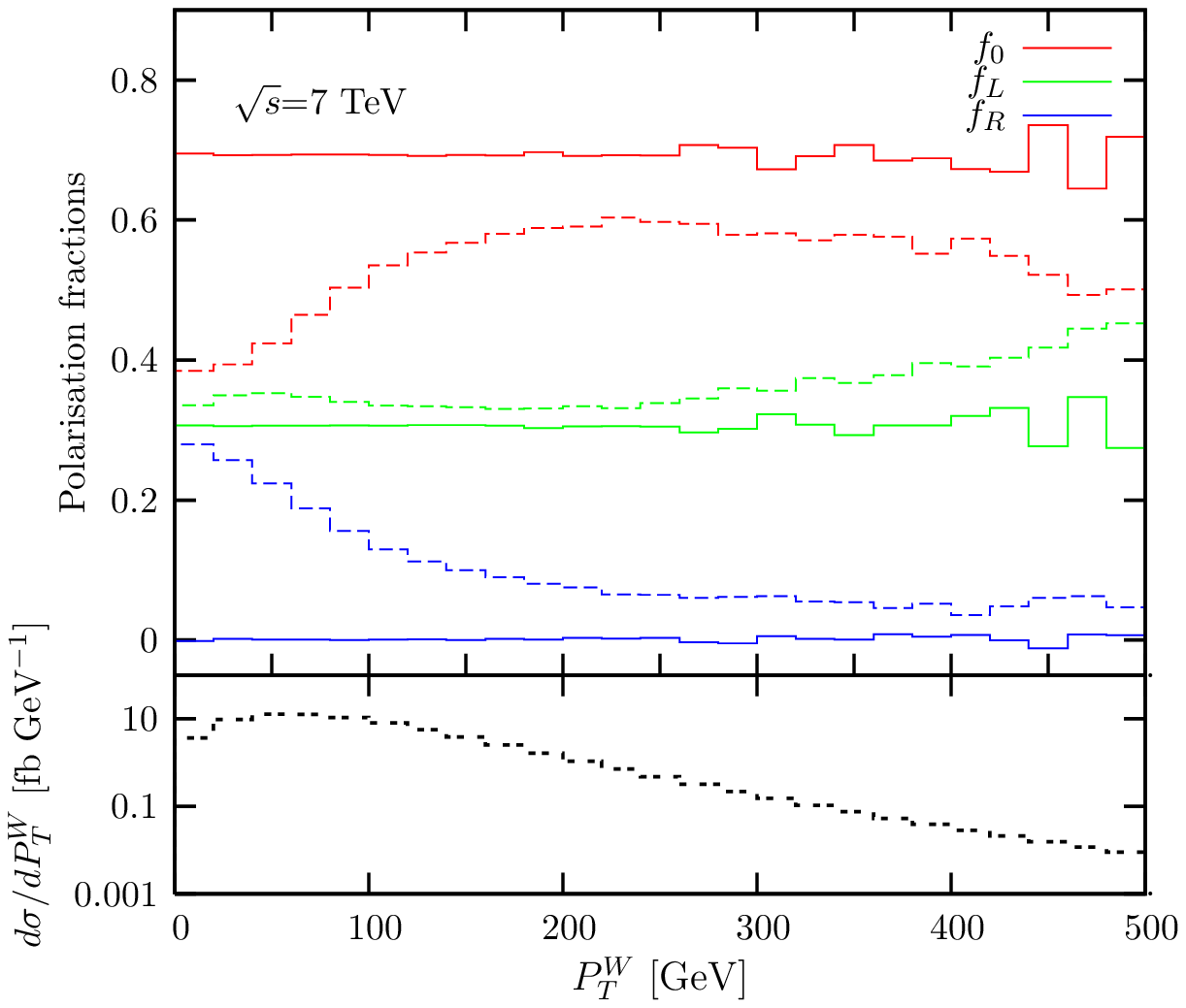}
 \caption{Polarisation fractions for $W^+$ from top decays with no imposed cuts
using the two angle definitions. Solid: top rest frame and dashed: lab frame.}
 \label{tnew}
 \end{figure}

In the extraction of polarisation fractions in order to be consistent with our analysis for  $W+$~jets we should of course 
use the same projections
with $\theta^*$ defined relative to the $W$ direction in the lab frame. The comparison
between the two definitions is shown in Fig.~\ref{tnew}, together with the distribution
for the $W$ $p_T$. For $W^-$ $f_R \Leftrightarrow  f_L$. It is clear from the plot that the polarisation fractions are highly sensitive to the definition of the angle and therefore frame dependent. The total polarisation fractions obtained by integrating over the whole phase space with no imposed cuts are different. Moreover, use of the lab frame in the definition also introduces a dependence of the fractions on the $p_T$ of the $W$ boson.

\section{$W$ bosons from other hard-scattering processes}

\begin{figure}[ht]
\centering
\subfigure[$W$ pair production (W$^-$ decays hadronically): 2048~fb]{
\includegraphics[trim=1cm 0 0 0,scale=0.45]{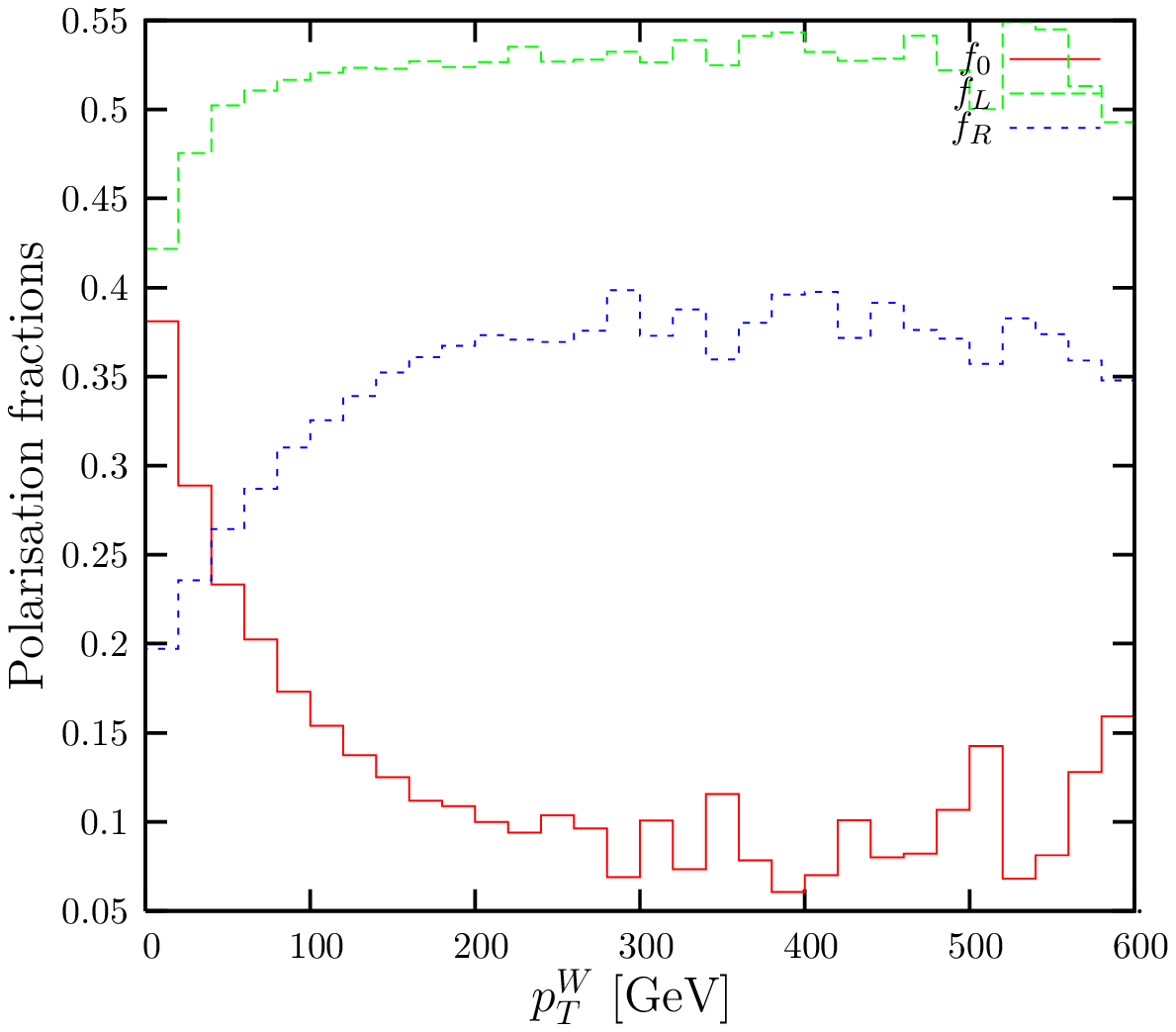}
\label{wpair}
}
\subfigure[$W +$~Higgs ($m_H=120$~GeV, Higgs decays to $b\bar{b}$): 53~fb]{
\includegraphics[trim=1.5cm 0 0 0,scale=0.45]{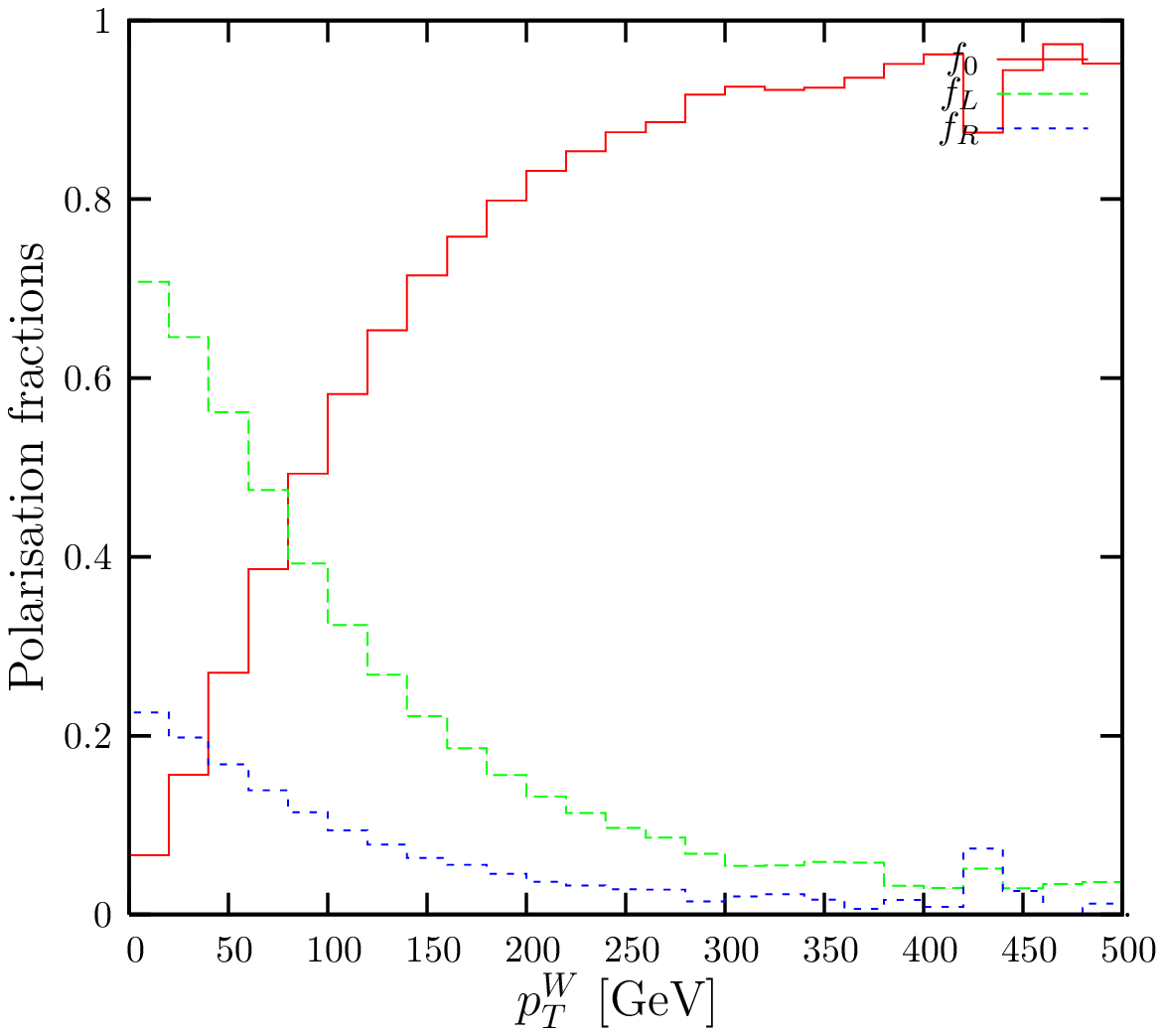}
\label{WH}
}
\subfigure[$W + Z$ ($Z$ decays hadronically to 3$\times d\bar{d}$): 622~fb]{
\includegraphics[trim=1cm 0 0 0,scale=0.45]{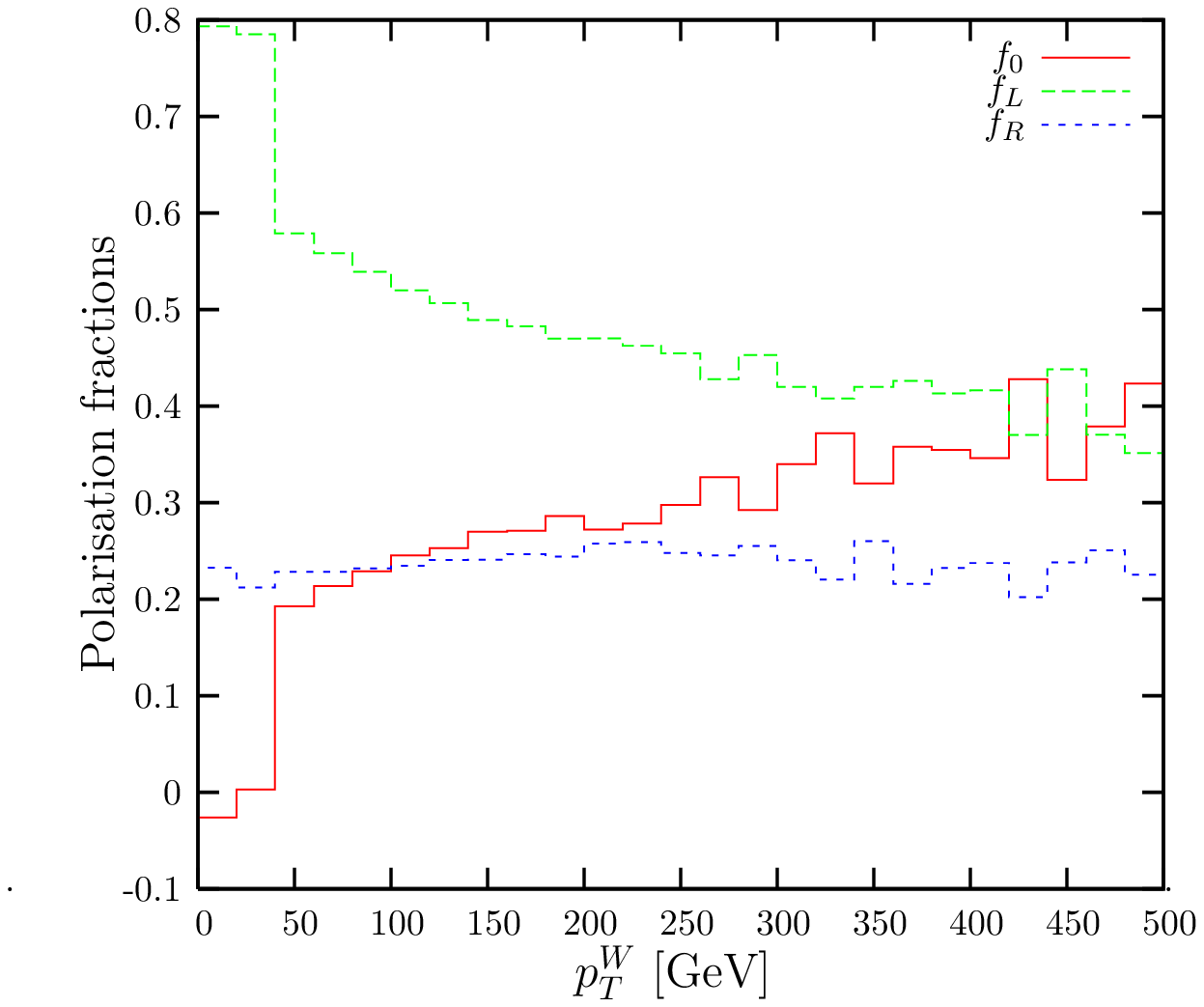}
\label{WZ}
}
\subfigure[Single top $t$-channel: 4067~fb ]{
\includegraphics[trim=1.2cm 0 0 0,scale=0.45]{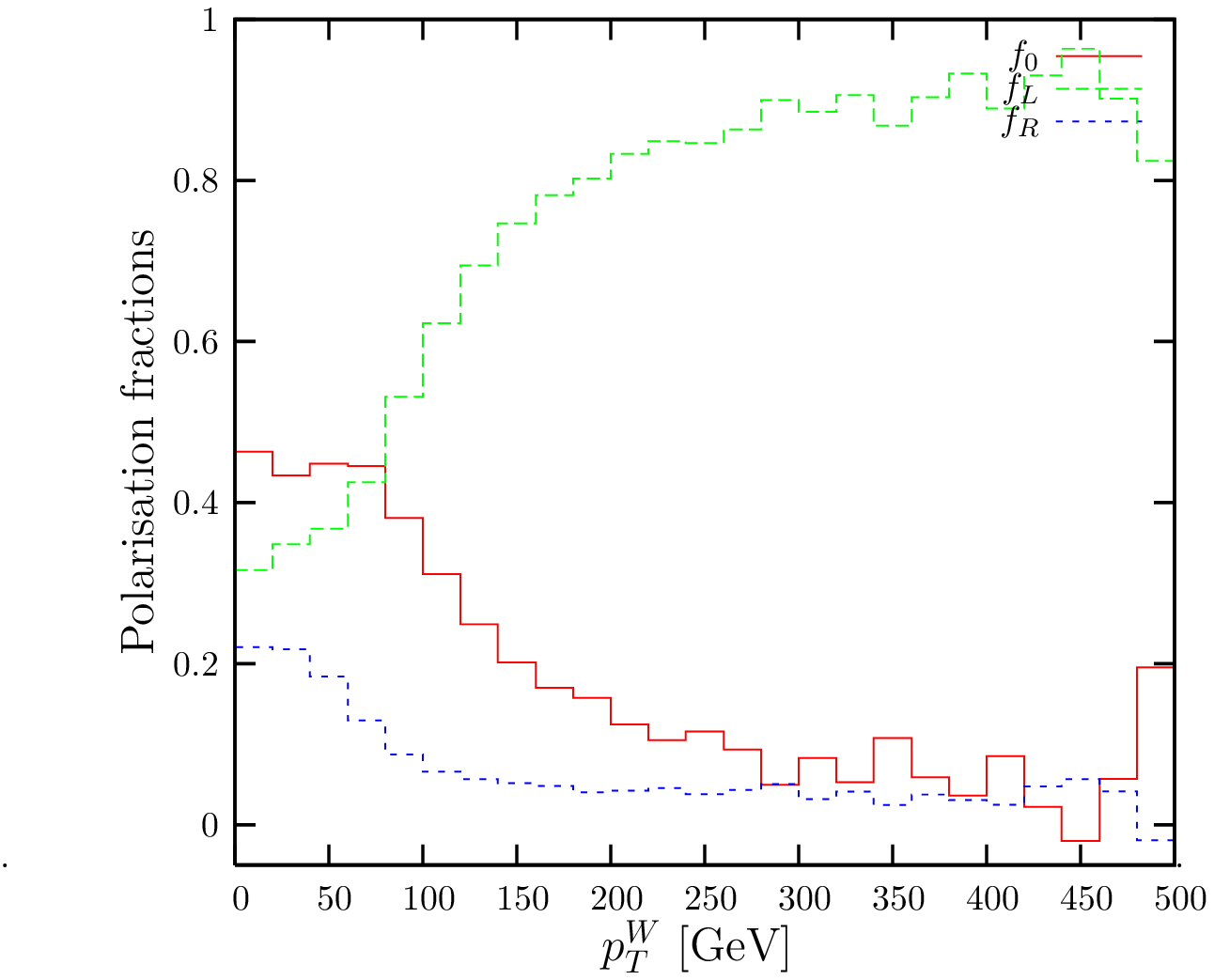}
\label{fig:subt}
}
\subfigure[Single top $s$-channel: 205~fb]{
\includegraphics[trim=1.2cm 0 0 0,scale=0.45]{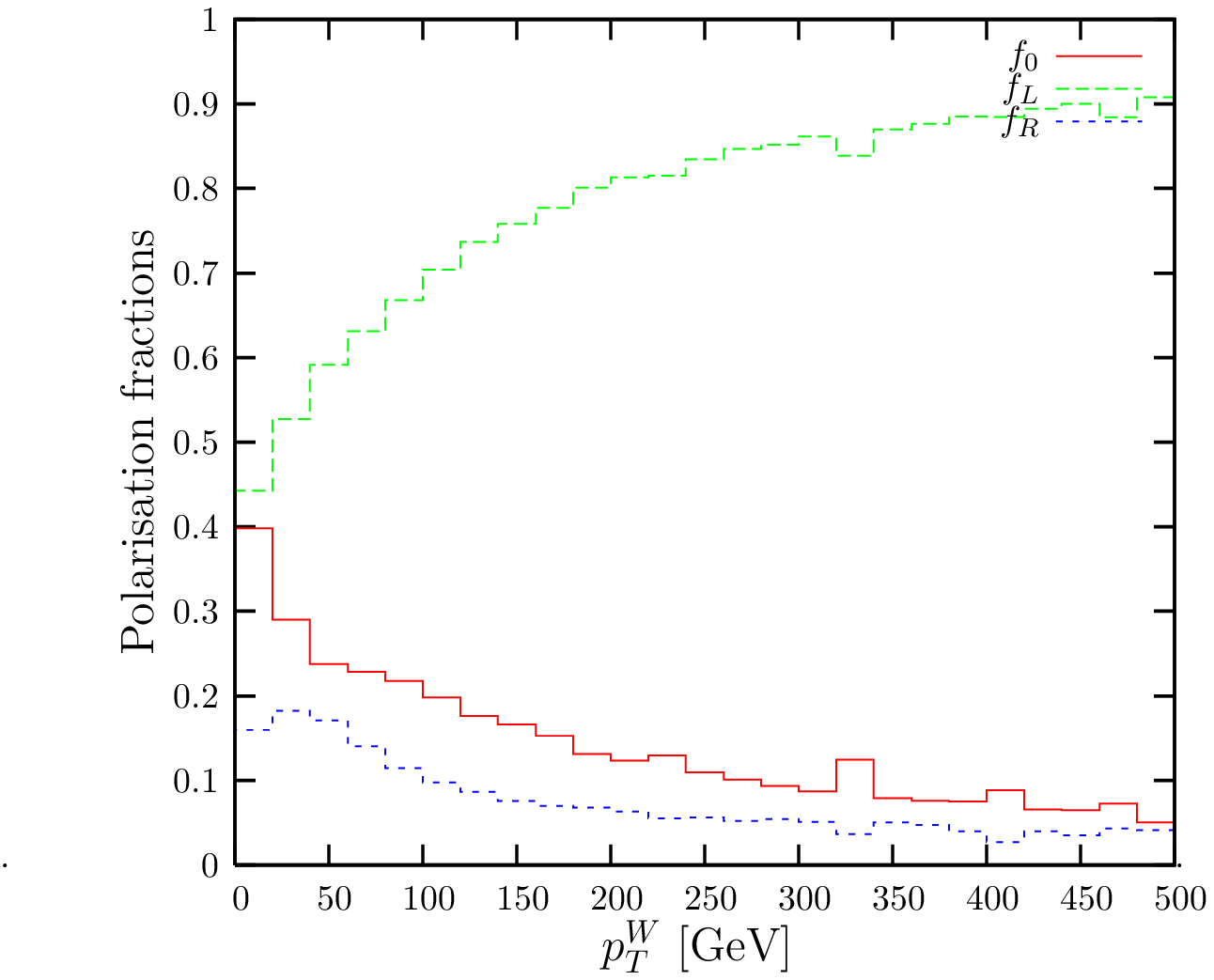}
\label{fig:subts}
}
\subfigure[$W$ pair from Higgs decay ($m_H=120$~GeV): 10.3$~$fb]{
\includegraphics[trim=1.2cm 0 0 0,scale=0.45]{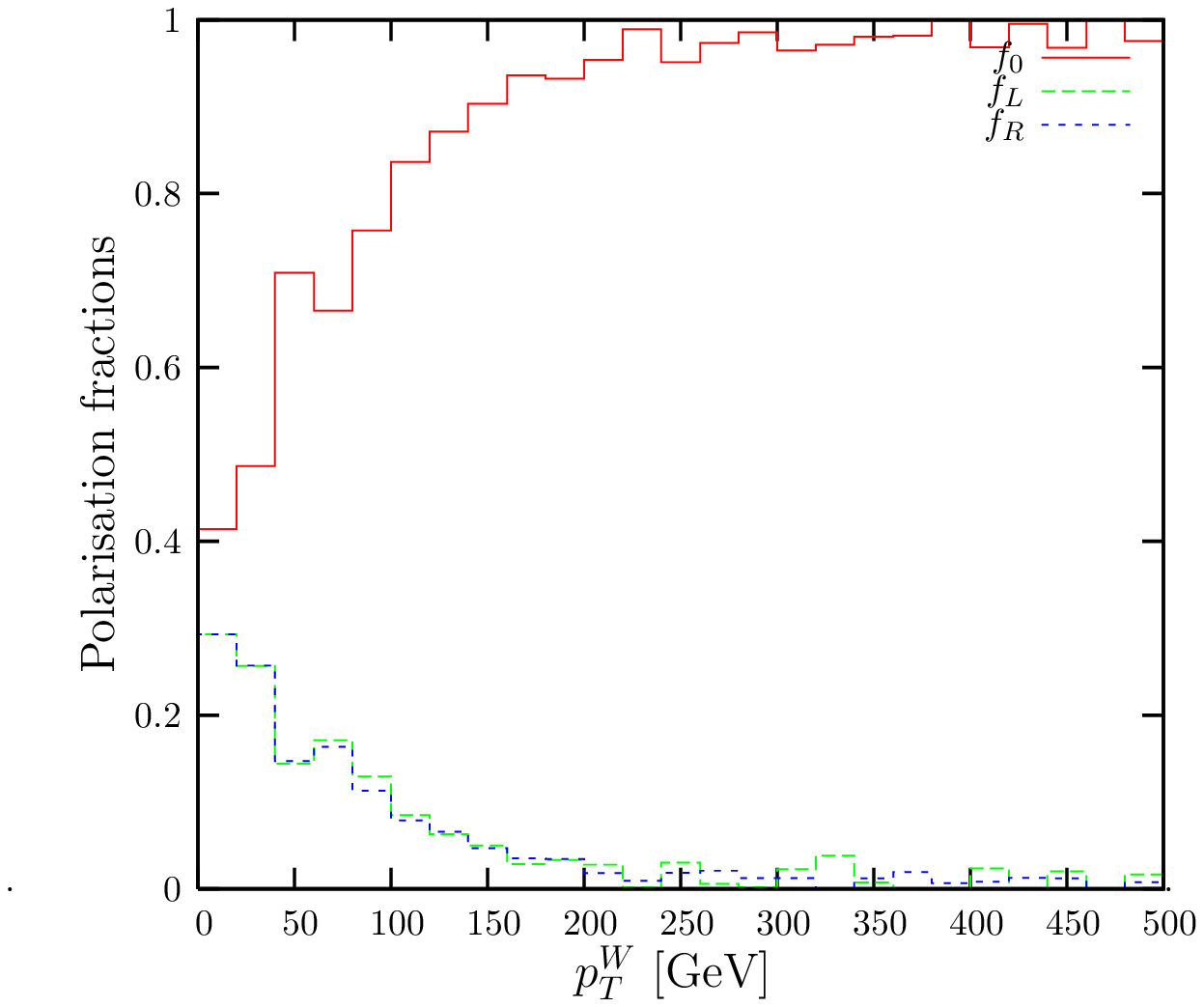}
\label{fig:subH}
}
\caption[Optional caption for list of figures]{Polarisation fractions for
different processes.}
\label{diff2}
\end{figure} 
Apart from QCD $W+$~jets and top pair production, other processes producing $W$ bosons
include: single top production, $W^+W^-$ and $WZ$ gauge boson pair production, $W$ plus
Higgs production and $H\to W^+W^-$ decay. In this section we compare the polarisation properties of these $W$ bosons 
with those from standard QCD $W+$~jets production. In each case we use the same projections as 
for  $W+$~jets defined in Section~\ref{sec:2} and select the available MCFM subprocesses where 
the $W^+$ decays to a positron and a neutrino and the other particles produced in the event ($W^-$, $H$, $Z$) decay 
hadronically so that for each event there is only one charged lepton present.

We first note that the polarisation fractions of the produced $W^+$ boson vary from process to
process as shown in the plots of Fig.~\ref{diff2}, for which no selection cuts have been applied. 
For each case we also give the LO positron channel subprocess cross section result obtained using MCFM as a 
guide to the relative importance of each process as a $W$ source.\footnote{Similar calculations can of course 
be made for $W^-$ production but here we only show the results for $W^+$.}

These processes are in general a subdominant source of $W$ bosons compared to QCD $W +$~jets but it is still 
 important to explore them both because they constitute a further set of backgrounds for New Physics searches 
and also because they are interesting  processes in their own right. The cross-section results at 
7~TeV and the total polarisation fractions are given in Table ~\ref{smtable}. 
In all cases the $W^+$ decays leptonically to a positron and a neutrino leading to missing transverse energy. 
In comparison with the distributions we note that the total 
fractions are more influenced by the fractions at low $p_T^W$ as in general the cross 
sections are rapidly decreasing functions at large $p_T^W$. The only cut imposed for the 
results shown in Table \ref{smtable} is a jet $p_T$ cut of 30~GeV for the $W+$~jets processes. 
As the cross sections for the non-QCD processes are much smaller, the experimental determination of 
the polarisation fractions could well be impeded by low statistics.  

\begin{table} [h]
\caption{Comparison of results for the different processes leading to $W^+$ production. }
\begin{center}
    \begin{tabular}{ | c | c | c | c | c |}
    \hline
   Process  & Cross section[fb] & $f_0$ & $f_L$ & $f_R$   \\ \hline
    $W+1$ jet($p^j_T>30$ GeV) & 6.11$\cdot 10^5$ & 0.20 & 0.56 & 0.24 \\ \hline
    $W+2$ jets($p^j_T>30$ GeV) & 2.15$\cdot 10^5$& 0.20 & 0.56 & 0.23\\ \hline
    $W+3$ jets($p^j_T>30$ GeV) & 0.74$\cdot 10^5$  & 0.21 & 0.56 & 0.23  \\ \hline
    $t\bar{t}$($\bar{t}\rightarrow bq\bar{q}$) &  1489 & 0.46 & 0.37 & 0.17 \\ \hline
    Single top($t$-channel) &  4067 & 0.42 & 0.43 & 0.15  \\ \hline
    Single top($s$-channel) &  205 & 0.24 & 0.61 & 0.14  \\ \hline
    $W+Z$($Z\rightarrow 3\times (d\bar{d})$)& 622 & 0.05 & 0.72 & 0.23 \\  \hline
    $W+H$($H\rightarrow b\bar{b}$) & 53 & 0.05 & 0.72 & 0.23 \\  \hline
    $W$ pair($W^-\rightarrow q\bar{q}$) & 2048 & 0.26 & 0.48 & 0.25  \\  \hline
    $W$ pair from $H$($m_H=120$ GeV, $H\rightarrow 4 l$) & 10.3 & 0.46  & 0.27 &  0.27 \\  \hline 
   \end{tabular}
\end{center}
\label{smtable}
\end{table} 

\begin{figure}[h]
\centering
\subfigure[]{
\includegraphics[trim=1cm 0 0 0,scale=0.53]{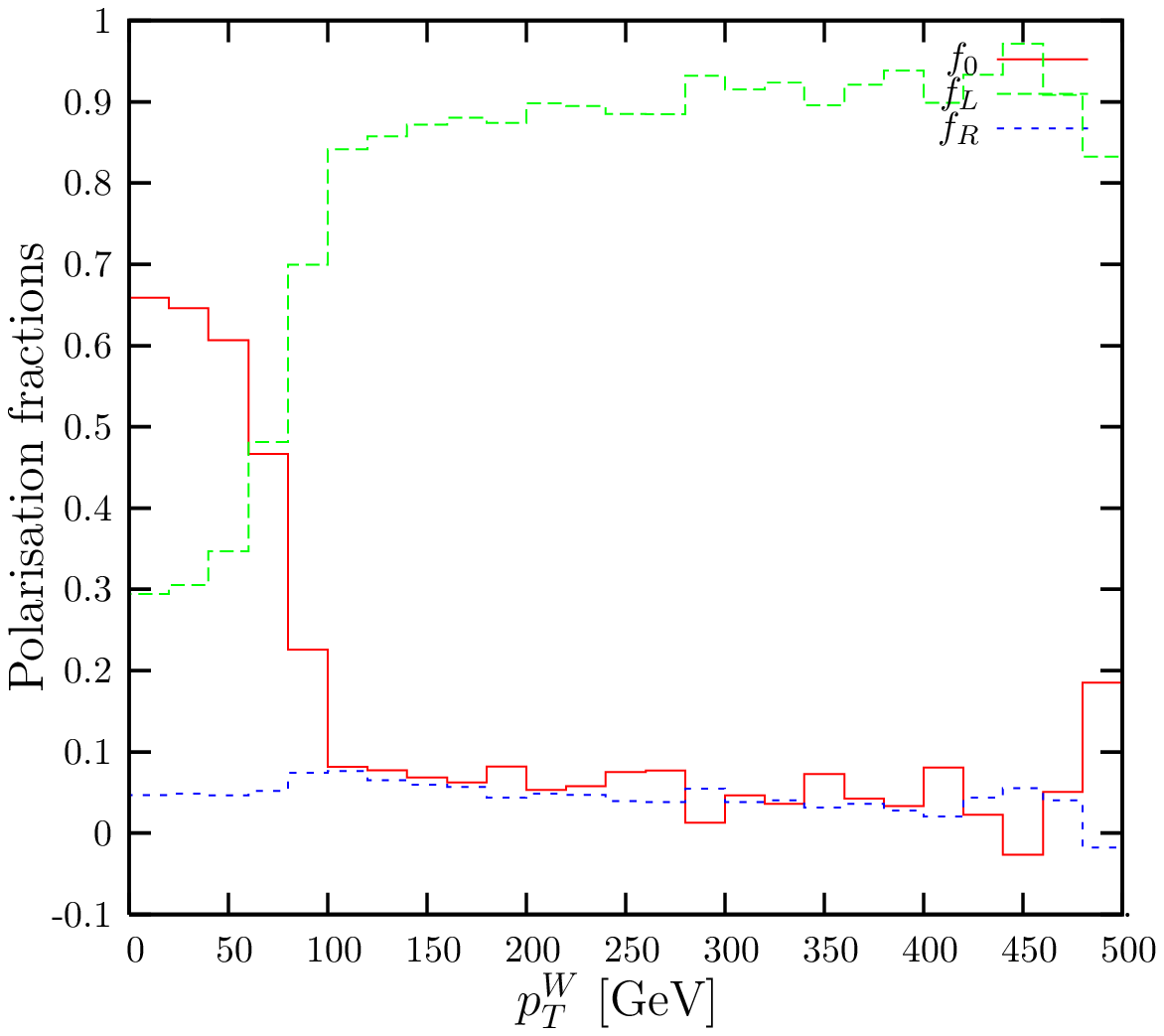}
}
\subfigure[]{
\includegraphics[trim=1cm 0 0 0,scale=0.53]{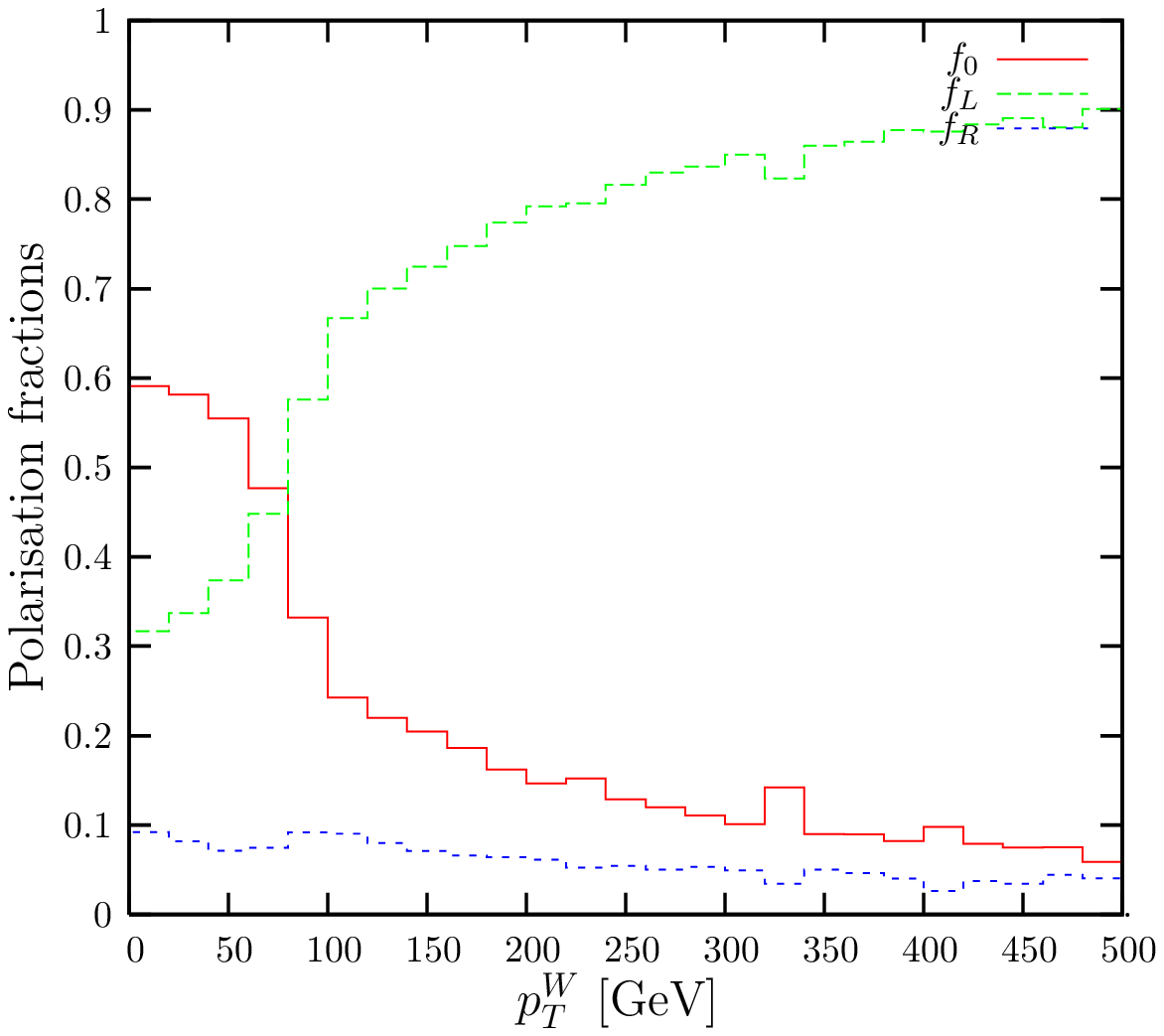}
}
\caption{Polarisation fractions for $W^+$ from single top a) $t$-channel and b) $s$-channel production using the 
direction of the $W$ in the top rest frame.}
\label{single}
\end{figure}

 For single top production we also show, in Fig.~\ref{single}, the result 
obtained using the projections defined in \cite{AguilarSaavedra:2006fy}. 
We note that the two definitions give results that coincide at high $p_T^W$, 
which is explained by the fact that the $W$ direction in the top rest frame and the
lab frame coincide for large $p_T^W$. At very high $p_T^W$, the 
left-handed $b-$quark is preferentially produced antiparallel to the top direction and 
the left-handed $W$ is produced parallel to the top by angular momentum conservation. 
The sum of the $W$ and $b$ momenta gives the top momentum and therefore the $W$ 
transverse momentum is on average larger than that of the top --- the top \lq sees' the $W$ boosted 
in the same direction as seen in the lab frame. 

It is clear from the plots and from Table~\ref{smtable} that the polarisation properties of $W$ bosons 
depend strongly on the production process.
One could try to explain the results based on the helicities of the
particles involved at scattering amplitude level, as attempted in Ref.~\cite{Bern:2011ie} for QCD $W+1$~jet production. 
However this rapidly becomes complicated and no clear conclusions can be drawn. 

One further step that could be taken is to try to locate the origin of the differences between a pair of processes. We
expect the difference between two processes to be partly caused by different kinematics and
partly by the physics of the underlying interactions. In order to disentangle the effect of kinematics from the 
interaction we can impose cuts on one process to mimick the kinematics of another
process.

\begin{figure}[h]
\begin{minipage}[b]{0.5\linewidth}
\centering
\includegraphics[trim=1.2cm 0 0 0,scale=0.6]{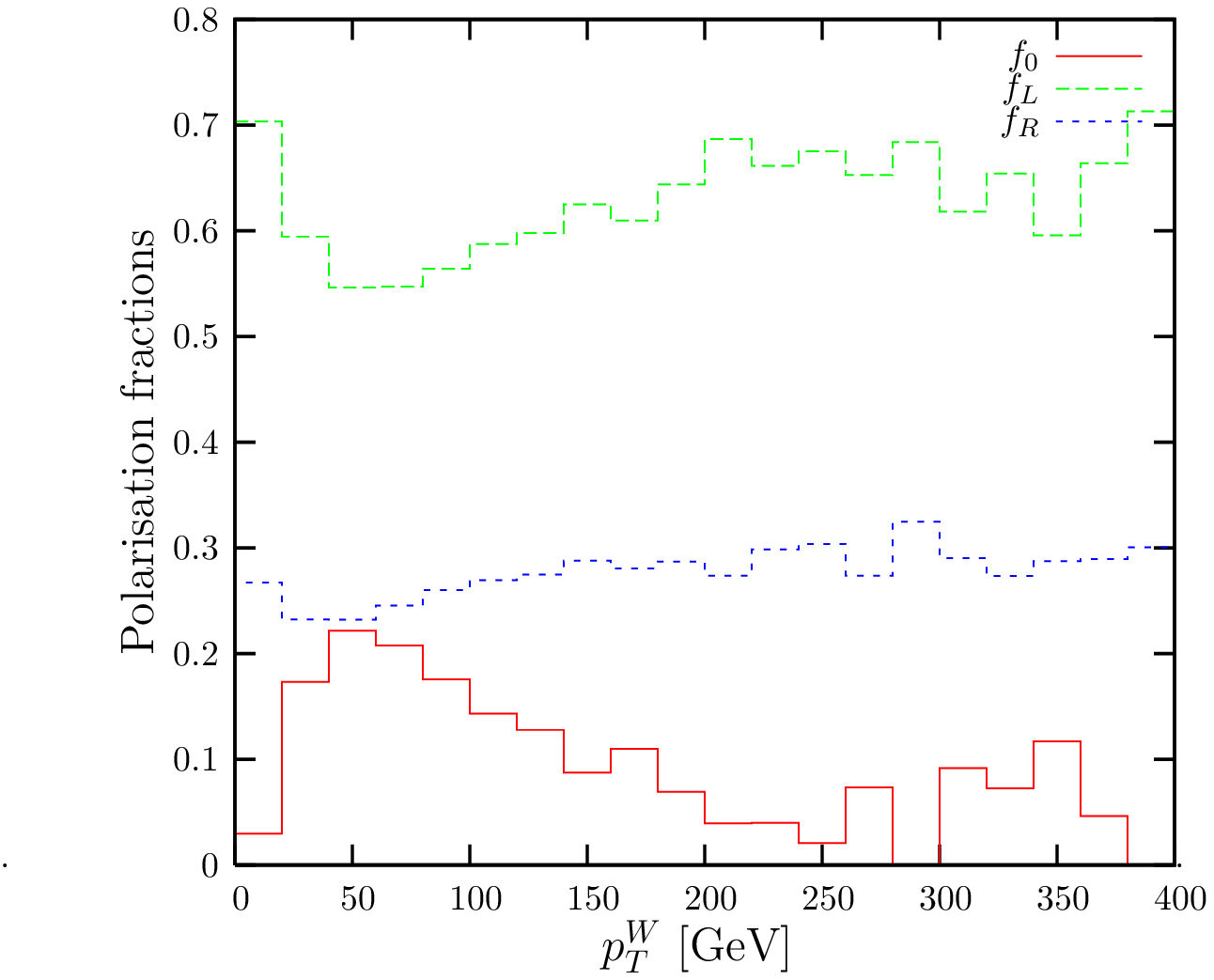}
\caption{Polarisation fractions for $W^++2$~jets production for the dijet mass of the jet pair constrained to be 
close to $M_Z$.}
\label{W2Zall}
\end{minipage}
 \hspace{0.5cm}
\begin{minipage}[b]{0.5\linewidth}
\centering
\includegraphics[trim=1.2cm 0 0 0,scale=0.6]{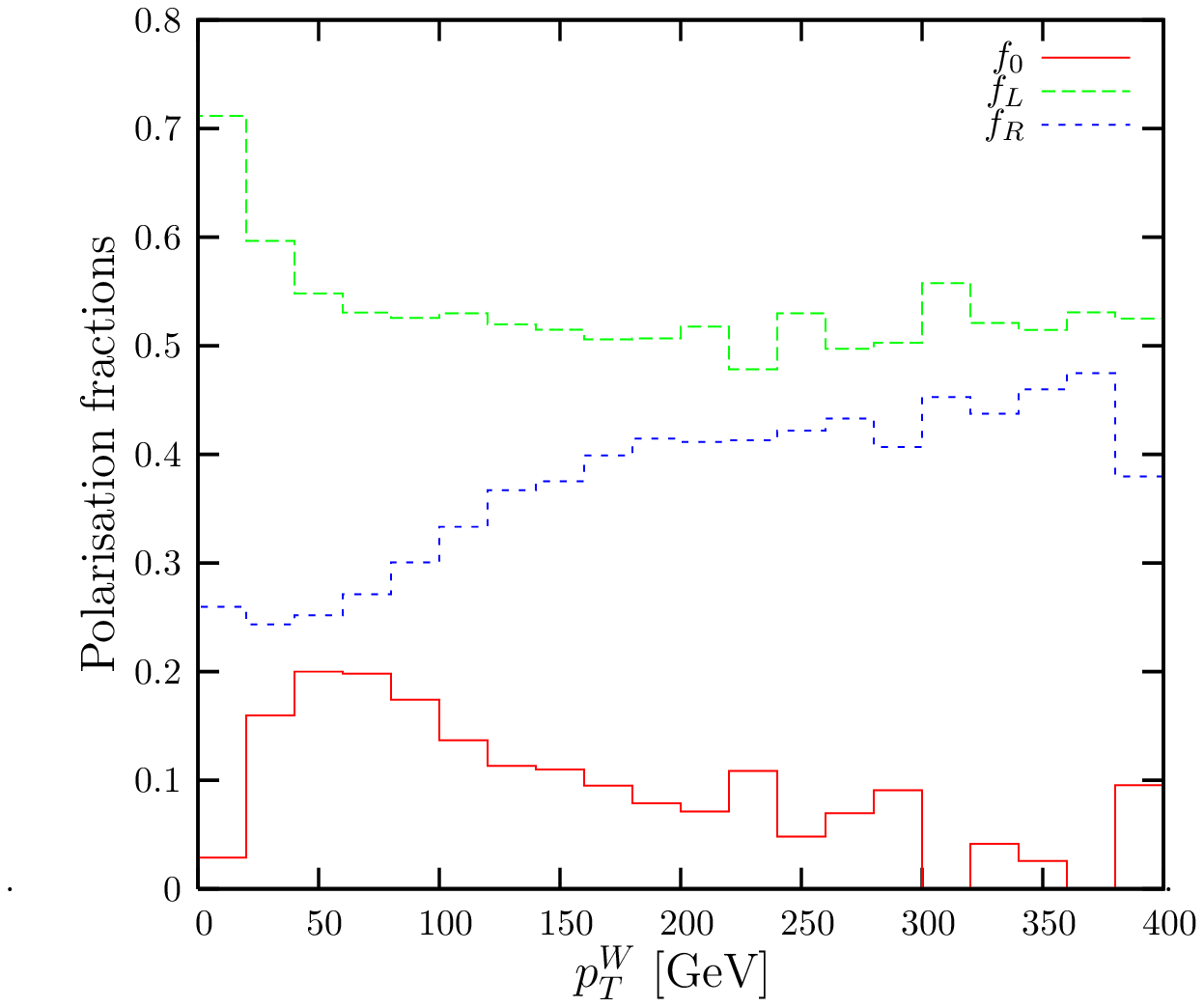}
\caption{Polarisation fractions for quark only contribution to $W+2$~jets production 
mimicking the kinematics of $W+Z$. }
\label{qqWZ}
\end{minipage}
\end{figure}

As an example we compare $W+Z$ associated production where the $Z$ decays hadronically 
with QCD $W+2$~jets production where we impose a cut on the jet pair invariant mass to force it to be very close to
the $Z$ mass. The effect of this constraint is shown in Fig.~\ref{W2Zall}. Of course we must bear 
in mind that in $W+Z$ production the $Z$ can be off-shell and also that a small admixture of photon events is included 
but we can check this by examining the
invariant mass of the two final state jets for this process. 
In the dijet mass differential cross section we can see the rapid increase around
the $Z$ mass pole. There is also an increase at small invariant masses due to the contribution 
of the virtual photon which is also included in the MCFM calculation. 
By selecting the appropriate subprocess in which the $Z$ decays to neutrinos and therefore no photon events 
can be included, we can examine the impact of the photon events in the polarisation properties of the $W$ boson. 
The shape is only modified at small transverse momentum, but the basic characteristics at high $p_T$ 
persist even though the crossing between $f_L$ and $f_0$ shifts to a higher $p_T$ value. 

Another consideration in this comparison is the dominant parton subprocess for the two processes. 
For $W+2$~jets production at the LHC quark-gluon scattering dominates over quark-quark scattering. 
On the other hand at LO $W+Z$ production originates only from quark-antiquark scattering. 
To extract more information we can decompose the $W+2$~jets cross section into contributions from quark 
and gluon scattering components by setting the gluon PDF to zero.
 The results for
this are shown in Fig.~\ref{qqWZ}. As the imposed constraints decrease the cross section dramatically, 
lack of statistics inevitably affects the results at high $p_T$. Comparing this to the result for $W+Z$,
we notice that the shapes remain quite different especially at high $p_T^W$. Therefore the polarisation behaviour here
originates mainly from the underlying interaction. This is consistent with the fact that for
$W+$~jets production with no initial state gluons the two outgoing jets are gluon jets of QCD origin. On
the other hand, in the $W+Z$ case the two jets are quark jets that originate from an electroweak interaction
that also distinguishes fermion helicities through the right-handed and left-handed couplings.

Also related to the underlying interaction in the process producing the $W$ boson is the 
similarity in the shapes of the results for $W+H$ associated production and $H\to W^+W^-$ Higgs decay. 
In both cases the $W$ boson is predominantly longitudinal at high $p_T$ which is related to 
the spin-0 nature of the Higgs boson involved in the interaction. The difference between the two is the 
absence of an asymmetry between left and right polarisation for $W$ from Higgs decay. 
The asymmetry observed in $WH$ production at small $p_T$ is related to the initial state 
quark--anti-quark distributions producing a preferentially left-handed $W$. 

As already stressed above, these are complicated processes and it is not straightforward to predict the
results in a simple analysis such as that performed in \cite{Bern:2011ie} for $W+1$~jet production. One
region for which we do have
some intuition is the region of high $W$ $p_T$ for $W+Z$, $W+H$ and single top 
production using a simple angular momentum conservation argument. For $W+H$ production, by
considering that only left-handed quarks are involved and that the Higgs is a
scalar, we expect the $W$ to be predominantly longitudinal. On the other hand, for $W+Z$ production 
we expect the $f_L$ and $f_R$ polarisation fractions to be the same. 
This can be seen by drawing angular momentum conservation diagrams with the momentum and spin vectors of each 
particle, similar to  those used in \cite{Bern:2011ie} for $W+1$~jet production. 

We also note that $f_0\rightarrow 0$ at large $p_T^W$ except for $W$ plus Higgs and $W+Z$. This is related to the equivalence theorem which states that at high energies the longitudinally polarised gauge boson states can be replaced by the corresponding Goldstone bosons which do not couple to light quarks. For the processes involving only light quarks this implies that $f_0\rightarrow 0$ at high energies.

\section{$Z$ boson polarisation}
The methods defined and employed for $W$ bosons can in principle also be used to measure the polarisation of 
$Z$ bosons produced at the LHC.
In the case of the $Z$ boson decaying to two charged leptons, the shape of the positively and negatively charged 
lepton distributions  cannot be used as a probe of polarisation in the same way as for the $W$ boson, because the $Z$ 
couples to {\it both} right- and left-handed fermions. 
Therefore the SM leptonic couplings need to be taken into account and the equivalent of Eq.~(\ref{diffeq}) for $Z$ decay 
to a pair of fermions is:
\begin{eqnarray}\nonumber
 \frac{1}{\sigma}\frac{d\sigma}{d\text{cos}\theta^*}&=&\frac{3}{8}\bigg(1+\text{cos}^2{\theta^*}-
\frac{2(c_L^2-c_R^2)}{(c_L^2+c_R^2)}\text{cos}\theta^*\bigg) f_L+\frac{3}{8}\bigg(1+\text{cos}^2{\theta^*}
+\frac{2(c_L^2-c_R^2)}{(c_L^2-c_R^2)}\text{cos}\theta^*\bigg)f_R\\
&+&\frac{3}{4}\text{sin}^2\theta^*f_0,
\label{diffeqZ}
\end{eqnarray} with $c_R$ and $c_L$ the right- and left-handed couplings of the fermion to the $Z$ and 
$\theta^*$ the angle measured in the $Z$ rest frame between the antiparticle
 and the $Z$ flight direction in the lab frame. Based on Eq.~(\ref{diffeqZ}) and similarly to $W$ boson production 
we can use appropriate projections, which are however in this case dependent on the couplings, 
to obtain  the polarisation fractions:
\begin{eqnarray}
f_0&=&2-5\langle \rm{cos}^2\theta^{*}\rangle,\\
f_L&=&-\frac{1}{2}-\frac{(c_L^2-c_R^2)}{(c_L^2+c_R^2)}\langle \rm{cos}\theta^*\rangle+\frac{5}{2}\langle \rm{cos}^2\theta^{*}\rangle, \\
f_R&=&-\frac{1}{2}+\frac{(c_L^2-c_R^2)}{(c_L^2+c_R^2)}\langle \rm{cos}\theta^*\rangle+\frac{5}{2}\langle \rm{cos}^2\theta^{*}\rangle.
\end{eqnarray} 
For the decay to neutrinos the projections are identical to those for $W$ bosons, since $c_R^{\nu}=0$. 
Using our MC generators we can use different decay channels to determine the polarisation fractions of $Z$ bosons.
In contrast to the case of $W$ bosons, for which $\theta^*$ cannot be extracted precisely due to the unreconstructed 
longitudinal $W$ momentum, for $Z$ production $\theta^*$ is unambiguously defined and reconstructed. 
Therefore there is no need to use the variable $L_p$ introduced earlier. This method can therefore be used directly to 
determine the polarisation of $Z$ bosons from different processes. Here we briefly comment on QCD $Z+$~jets production and 
$Z$ bosons produced in association with $W$.

\subsection{$Z$ plus jets}
\begin{figure}[h]
\centering
\includegraphics[trim=1.2cm 0 0 0,scale=0.6]{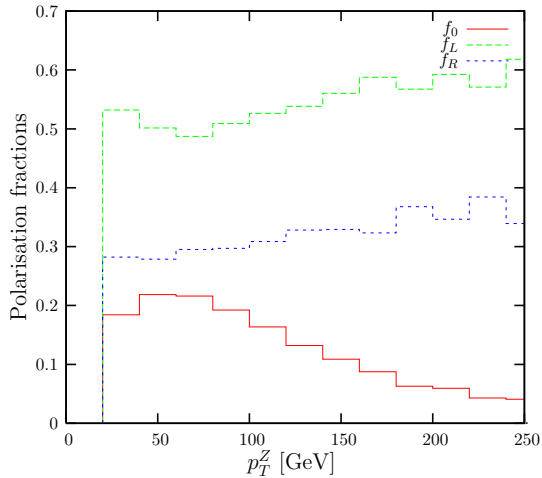}
\caption{$Z$ polarisation fractions from the electron decay channel.}
\label{nuel}
\end{figure}

 The polarisation results are shown in Fig.~\ref{nuel} for $Z+1$~jet at 7 TeV with a jet $p_T$ cut of 30~GeV. 
Evidently $Z$ bosons are also predominantly left-handed at non-zero transverse momentum. 
Comparing the results for $Z$ to those for $W+1$~jet  we note that while the $Z$ is also predominantly left-handed, 
the exact values of the fractions differ, as they arise from a combination of the different quark flavour combinations 
producing the $Z$ boson and the fact that the $Z$ couples to both left- and right-handed quarks. 
A complication that arises in $Z$ boson studies is the small admixture of photon events. 
In terms of the polarisation fraction expressions given above, these apply only to pure $Z$ exchange. 
Experimentally this is limited by introducing a cut constraining the lepton pair invariant mass to be close to the $Z$ mass. 
In practice, for example in the CMS $Z$ boson analysis~\cite{Chatrchyan:2011wt}, this is constrained to lie 
between 60 and 120 GeV. Using MCFM we can allow for the inclusion of photon events in this region of masses, 
obtaining the LO cross section for $Z+1$~jet production with a jet $p_T$ cut of 30~GeV of 105.9~pb when these are 
included and 103.3~pb when they are excluded. Given the small percentage difference and the appropriate use of 
MC generators a determination of the polarisation fractions of $Z$ from the data sample should therefore be possible.

\subsection{$W$ plus $Z$}
\begin{figure}[h]
\centering
\subfigure[]{
\includegraphics[trim=1.2cm 0 0 0,scale=0.56]{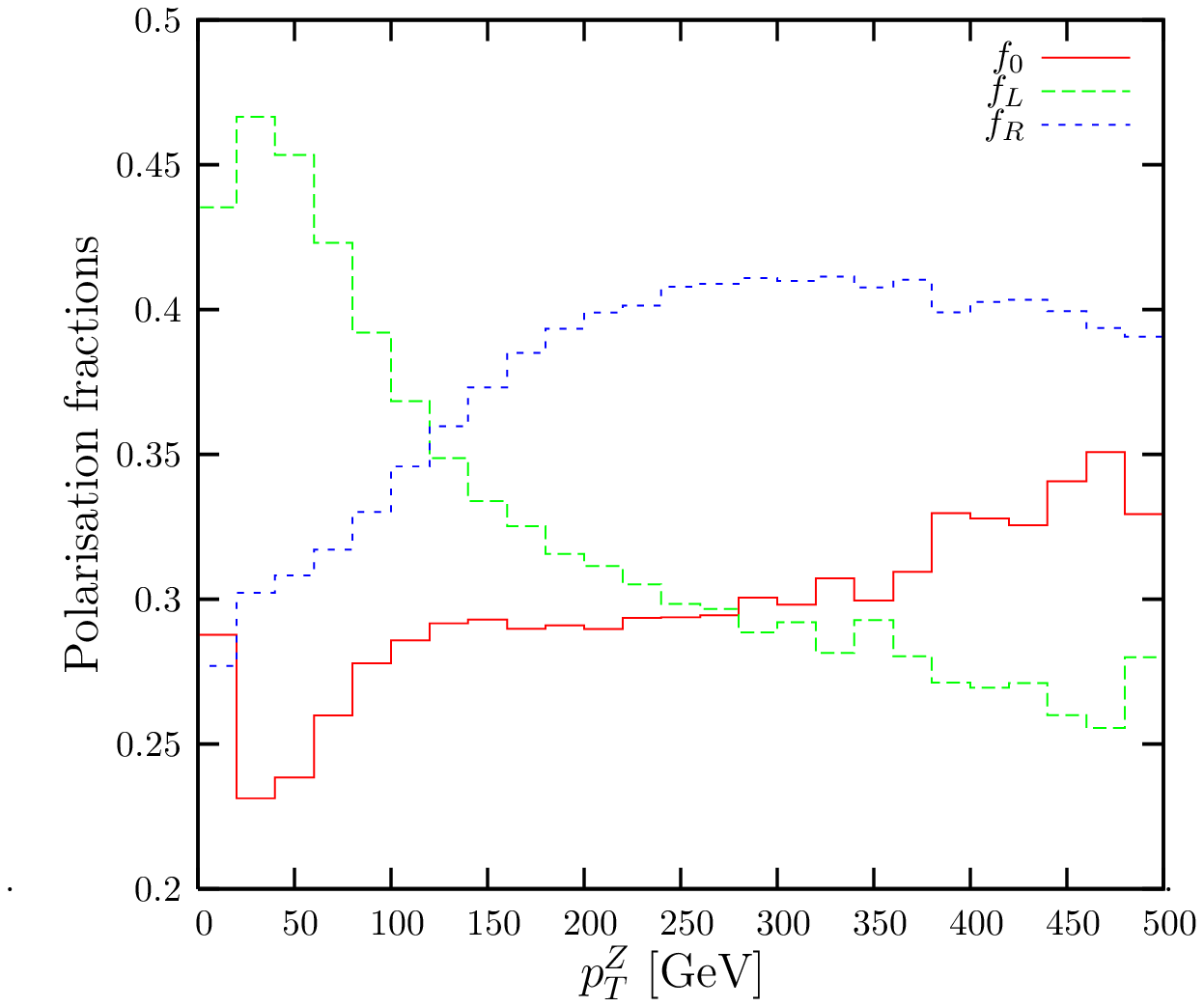}
}
\subfigure[]{
\includegraphics[trim=1.2cm 0 0 0,scale=0.56]{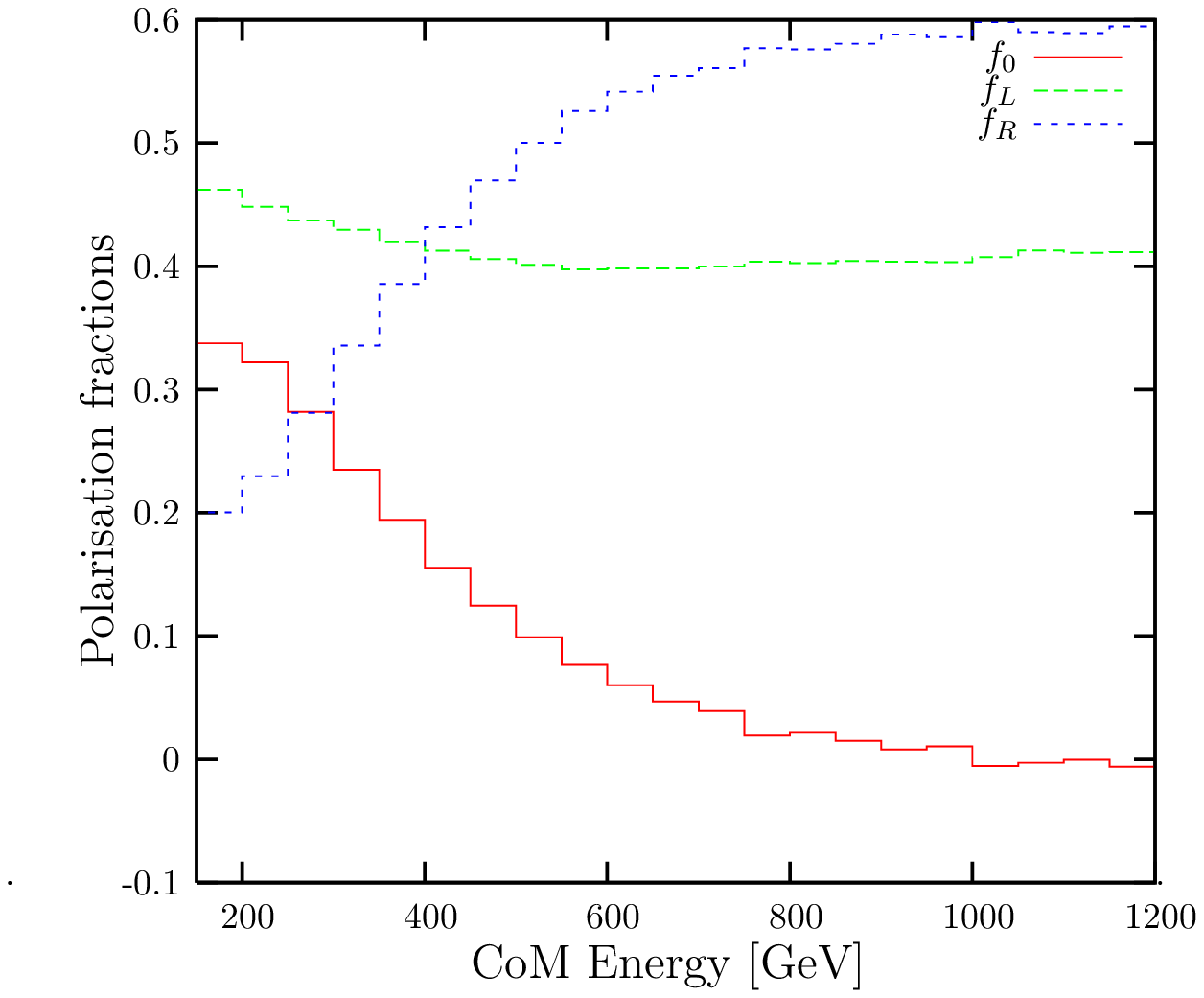}
}
\caption{Polarisation fractions for $Z$ bosons from $W^+ + Z$ production as a function of a) the $Z$ $p_T$ and b) 
the centre-of-mass energy. }
\label{WZpol}
\end{figure}

Similarly to the measurement of the polarisation of the $W$ in $WZ$ production, 
we can extract the polarisation of the $Z$ in the same process. For this we use the 
MCFM subprocess for which the $Z$ decays to neutrinos to avoid the small admixture of photon events. 
The results are shown in Fig.~\ref{WZpol} as a function of the $Z$ transverse momentum and the centre-of-mass energy. 
We note the increase in $f_0$ at high $p_T$. This can also be explored by considering the polarisation fractions 
as a function of the centre-of-mass energy from which we see that $f_0$ falls to zero at high energies. 
One might expect that high $p_T$ corresponds to high centre-of-mass energy and therefore that 
the limiting values should coincide. The fact that this is not observed to be the case 
can be investigated by setting a cut on the centre-of-mass energy and computing the $p_T$ distribution, and {\it vice versa}. 
Setting a cut $E_c$ on the centre-of-mass energy results in modified polarisation fractions at low $p_T$ 
while the fractions remain the same for $p_T> E_c/2$.  In contrast, imposing a cut $E_c/2$ on the $p_T$ forces the 
centre-of-mass energy to be larger than $E_c$ but also modifies the polarisation fractions at any given energy 
above $E_c$. These observations for the centre-of-mass energy and transverse momentum can explain the different 
shapes of the polarisation fraction distributions and the more complex connection between Figs.~\ref{WZpol} a) and b). 
Considering the differential distributions for $p_T^Z$ and centre-of-mass energy also helps us verify that 
even though the shapes are different they result as expected in the same total polarisation fractions 
when integrated over the whole phase space. Moreover we have checked that, as expected, 
when processes allowing photons are selected, at very low lepton pair invariant masses where photon events dominate, 
it can be seen that $f_0$ vanishes, as photons can only be transversely polarised.

Another observation that can be made by computing the polarisation fractions is the 
different behaviour of the polarisation of $Z$ in $W^++Z$ compared to $W^-+Z$. This is not unexpected, 
as in one case the $Z$ is produced from a $u\bar{u}$ pair and in the other from a $d\bar{d}$ pair. The different 
 right- and left-handed couplings to these fermions then leads to different polarisation fractions.

\section{Conclusions} 
We have studied the angular distributions of $W$ boson decay products to extract information on the 
corresponding polarisation fractions. We have seen that $W$ bosons  produced in association with QCD jets at 
non-zero transverse momentum are preferentially left-landed at the LHC. This leads to asymmetries between the 
charged lepton and neutrino transverse momentum distributions. The dependence of the angular distributions 
has been studied for different selection cuts and these were found to change rapidly on the introduction of 
cuts and therefore experimentally it will be necessary to correct for the cuts before extracting polarisation fraction information. 

We have compared the polarisation of $W$ bosons produced with QCD jets to $W$ bosons from top pair production and decay,
 calculating the polarisation fractions in the same frame and comparing the shape of different observable distributions 
and the asymmetry between the lepton transverse momentum and the missing transverse energy. 
We have also compared the polarisation fractions obtained in two different frames, to show that the polarisation fractions 
are strongly frame dependent. We remark that the $f_0=0.7$, $f_L=0.3$ fractions often mentioned in the literature 
are valid only when defining the relevant angle in the top rest frame.
  
We have used the same procedure for other $W$ producing processes at the LHC to study the polarisation of $W$ bosons 
as a function of the $W$ transverse momentum. Other processes have lower cross sections and measurements of the 
polarisation properties could well be impeded by low statistics, at least at present. 
However with increasing LHC luminosity, it should in the near future become possible to extract the $W$ polarisation fractions
 in the same way as has been done by the CMS and ATLAS collaborations for $W+$~jets production. 
Similarly, we expect the measurement of the $Z$ polarisation to be feasible at the LHC and have presented 
the relevant results. 

Comparing different $W$ processes with very different polarisation results we note that the origin of the 
difference is related to the underlying physics of the interaction and the helicities of the other particles involved. 
Therefore a study of the polarisation properties can be used in conjunction with kinematics to distinguish between 
different sources of $W$ bosons. This is also helpful for New Physics searches where new interactions might give 
different polarisation fractions and can therefore be used as a handle to disentangle the signal from the SM background.

\acknowledgments{E.V. acknowledges financial support from the UK Science and Technology Facilities Council. We thank Lance Dixon and Marco Peruzzi for useful discussions. }

\end{document}